\documentclass[preprint,showpacs,showkeys,amsmath,amssymb,eqsecnum,tightenlines]{revtex4}

\usepackage{dcolumn}
\usepackage{graphicx}

\usepackage{latexsym}
\usepackage{axodraw}

\newcommand{\cs}{ {Callan--Symanzik} }

\newcommand{\BPHZ}{{BPHZ}}
\renewcommand{\L}{\mathcal{L}}  
\newcommand{\Z}{\mathit{Z}}
\newcommand{\F}{{\mathcal F}}

\newcommand{\sm}{S-matrix }

\newcommand{\Legendre}{$ {Legendre}$ }

\newcommand{\phif}{{\phi}^4 ~model}
\newcommand{\N}{{\mathcal N}}
\newcommand{\gW}{{\mathcal W}}

\newcommand{\QAP}{ {quantum action principle~}}

\newcommand{\lWI}{local  {Ward} identity~}

\newcommand{\gWI}{global  {Ward} identity~}

\newcommand{\dx}{\!\!{\rm d}^4x\,}
\def\dxx#1{{\rm d}^4x_{#1}\,}

\newcommand{\sn}{S_{n}(p_{1},p_{2}, \cdots,p_{n}) }
\newcommand{\sna}{S_{A,n}(p_{1},p_{2}, \cdots,p_{n}) }
\newcommand{\snaa}{\int\!\prod_{i=1}^{n}\dxx{i}\,
    e^{i\sum_{j=1}^{n}{p_{j}\cdot x_{j}}}\,G_{A,n}}
    
\newcommand{\fur}{\int\!\!\prod_{i=1}^{n}\dxx{i}\,
    e^{i\sum_{j=1}^{n}{p_{j}\cdot x_{j}}}\,}            
\newcommand{\gn}{G_{n}(x_{1},x_{2}, \cdots,x_{n}) }
\newcommand{\gna}{G_{A,n}(x_{1},x_{2}, \cdots,x_{n}) }
\newcommand{\gnaa}{(i r^{-\half})^n
 \prod_{i=1}^{n}(\Box_{x_i}+m^{2})\,G_{n}(x_{1},x_{2}, \cdots,x_{n})}
\newcommand{\prodn}{\prod_{i=1}^{n}(\Box_{x_i}+m^{2})} 
 \newcommand{\prodl}{\prod_{i=1,i\neq l}^{n}(\Box_{x_i}+m^{2})}
\newcommand{\gnp}{G_{n}(p_{1},p_{2}, \cdots,p_{n}) }
\newcommand{\summ}{\sum_{i=1}^{n}\left(1+p_{i}\partial_{p_{i}}\right)}
\newcommand{\sumx}{\sum_{i=1}^{n}\left(1+x_{i}\partial_{x_{i}}\right)}

\newcommand{\cp}
{\{p_{i}\mid p_{i}^2 = m^{2};~{p_{i}}^0 > 0; ~i=1,2 \cdots n\}}
     
\newcommand{\half}{{\textstyle \frac{1}{2}}}

\newcommand{\delk}
{\left[\int\!\!{\rm d}^4x\,\half\phi(x)\Box\phi(x)\right]_4}
\newcommand{\delm}{\left[\int\!\!{\rm d}^4x\,\half\phi^2(x)\right]_4}
\newcommand{\deli}{\left[\int\!\!{\rm d}^4x\, {\frac {1}  {4!} }{\phi^4(x)}\right]_4} 
\newcommand{\delmm}{\left[\int\!\!{\rm d}^4x\, \half\phi^2(x)\right]_2}

\newcommand{\Del}[1]{{\Delta}_{#1}}

\def\pr{\partial}
\newcommand{\parr}[1]{{#1}{\partial}_{#1}}
\newcommand{\parl}{\partial_{\lambda}}
\newcommand{\pmu}{\parr{m}+\parr{\mu}}
\newcommand{\pmmu}{\parr{m^2}+\parr{\mu^2}}

\newcommand{\eq}{\begin{equation}}
\newcommand{\en}{\end{equation}}
\newcommand{\eqa}{\begin{eqnarray}}
\newcommand{\ena}{\end{eqnarray}}

\def\l{\langle}
\def\r{\rangle}

\newcommand{\mn}{\mu\nu}
\newcommand{\eps}{\varepsilon}

\newcommand{\ds}[1]{$#1$}

\newcommand{\Imn}{I^{(n)}_m}
\newcommand{\Iln}{I^{(n)}_l}

\newcommand{\Ion}{I^{(n)}_1}
\newcommand{\Idn}{I^{(n)}_\lambda}
\newcommand{\Ikn}{I^{(n)}_k}
\newcommand{\Itn}{I^{(n)}_2}

\newcommand{\zn}{z^{(n)}}
\newcommand{\an}{a^{(n)}}
\newcommand{\rn}{\rho^{(n)}}
\newcommand{\tzn}{\tilde{z}^{(n)}}
\newcommand{\lzn}{z^{(n)}_\lambda}

\newcommand{\Lren}{\L_{\rm ren}}

\def\Ga{\Gamma}
\newcommand{\Gren}{\Gamma_{\rm ren}}
\newcommand{\Grenk}{\Gamma_{\rm ren}^k}

\newcommand{\fdq}[2]{\frac{\delta #1}{\delta #2}}

\newcommand{\pdq}[2]{\frac{\partial #1}{\partial #2}}
\newcommand{\psdq}[1]{\frac{\partial }{\partial #1}}

\def\pslash#1{{\setbox0=\hbox{$#1$}
  \rlap{\ifdim\wd0>.7em\kern.22\wd0\else\kern.1\wd0\fi /}#1}}

\begin{document}

\setlength{\unitlength}{1mm}  

\footskip 50pt

\vspace{1.0cm}

\title{Conformal Transformations of S-Matrix in Scalar Field Theory
}
\author{Yong Zhang}
\email{yong@itp.uni-leipzig.de }
\altaffiliation{supported by Graduiertenkolleg
  ``Quantenfeldtheorie: Mathematische Struktur und physikalische
Anwendungen", University Leipzig. }
\affiliation{Institut f{\"u}r Theoretische Physik, Universit{\"a}t Leipzig,
Augustusplatz 10/11, D - 04109 Leipzig, Germany}

\begin{abstract}

In this paper, three methods for describing the conformal 
transformations of the S-matrix in quantum field theory are 
proposed. They are illustrated by applying the algebraic 
renormalization procedure to the quantum scalar field theory, 
defined by the LSZ reduction mechanism in the BPHZ renormalization 
scheme. Central results are shown to be independent of scheme 
choices and derived to all orders in loop expansions. Firstly, 
the local Callan--Symanzik equation is constructed, in which the 
insertion of the trace of the energy-momentum tensor is related 
to the beta function and the anomalous dimension. With this 
result, the Ward identities for the conformal transformations 
of the Green functions are derived. Then the conformal 
transformations of the S-matrix defined by the LSZ reduction 
procedure are calculated. Secondly, the conformal transformations 
of the S-matrix in the functional formalism are related to 
charge constructions. The commutators between the charges and 
the S-matrix operator are written in a compact way to represent the 
conformal transformations of the S-matrix. Lastly, the massive 
scalar field theory with local coupling is introduced in order to 
control breaking of the conformal invariance further. The 
conformal transformations of the S-matrix with local coupling 
are calculated.

\end{abstract}


\maketitle

\section{Introduction}

The S-matrix plays a fundamental role in quantum field theory. It is
always used to construct the cross section, which can be checked
by measurements of scattering experiments. On the other hand, it is also 
one way of defining quantum field theory. For example, in the Epstein--Glaser 
scheme, see \cite{epstein}, the locality and the unitarity of the S-matrix with 
local coupling determine the whole theory. Furthermore, symmetries of 
the S-matrix have been studied on an abstract level \cite{coleman, hagg}.

In this paper, we study the conformal transformations of the S-matrix in four
dimensional flat space-time. When the mass of particles can be neglected in 
high energy experiments, the theory could be regarded as conformally invariant
at least in the classical approximation. Hence, solving
problems of this type is helpful to simplify calculations or to find 
some identities in phenomenological physics. At the abstract level, 
it also may improve knowledge 
on how to control anomalies or breakings in quantum field theory.

The conformal transformations consist of the  {Poincar\'e} transformations,
the dilatation transformation and the special conformal transformation.
In all versions of ordinary quantum field theory the S-matrix has to
be  {Poincar\'e} invariant, which has been verified in all
experiments until now. But generally, the S-matrix is not invariant under 
the dilatation transformation and the special conformal transformation 
even if the corresponding classical theory is conformally invariant. 
Some research has been carried out on the breaking of the conformal invariance. 
For example, in the massless $\phi^4$ theory
constructed by a nonperturbative approach in \cite{Zimmer1},
the dilatation transformation of the S-matrix is given by
\eq
\sum_{i=1}^{n}\left(1+p_{i}\partial_{p_{i}}\right)S_{n}
=\beta_{\lambda}\partial_{\lambda}S_{n},
\en
where $\gamma$, the anomalous dimension, does not contribute.

We will treat our problem in the approach of the
 {algebraic renormalization} introduced in \cite{brs1,brs2}, also see 
\cite{sibold1, piguet, schweda}. It is based on the quantum action principle, relating
differentiation (or variation) on parameters (or fields) to local insertions
\cite{Low1, Low2, Lam1, Lam2}. There are two guiding arguments 
about  {the quantum action principle}. Firstly, it is independent of regularization 
schemes and renormalization procedures.  Secondly, the perturbative \QAP is satisfied 
to all orders of $\hbar$. Hence they provide strong power for
the algebraic renormalization procedure. Furthermore, the algebra 
of the global (local)  {Ward} identity operators 
(or the cohomology of the  {Slavnov}--{Taylor} identity operator) can be realized
by the differential (variational) operators.  By imposing  them as constraints
on local insertions, the global (local)  {Ward} identities 
(or the  {Slavnov}--{Taylor} identity) can be constructed to all orders in
loop expansions.

The global  {Ward} identity for the conformal transformation of the vertex functional
$\Gamma[\phi]$ can be defined respectively by
\eq
\label{gwic}
\mathcal{W}^{i}\Gamma[\phi]:=\int \dx
\delta^{i}\phi(x){\fdq{\Gamma[\phi]}{\phi(x)}},\qquad\qquad i=T,L,D,K,
\en
where the symbol $\mathit{T}$ denotes translation, the symbol $\mathit{L}$ denotes 
 {Lorentz} rotations, the symbol $\mathit{D}$ denotes  dilatation and 
the symbol $\mathit{K}$ denotes special conformal transformation. 
If we redefine the global  {Ward} identity operator
$\mathcal{W}^{i}$ by $-\mathcal{W}^{i}$, the new global  {Ward} identity operator
exactly forms a representation of the conformal algebra.
The corresponding local  {Ward} identity operator is given by
\eq
\mathbf{w}^i(x)=\delta^{i}\phi(x){\frac{\delta}{\delta\phi(x)}},
\qquad\qquad i=T,L,D,K.
\en
It is not unique since we can add total derivatives
without changing the global  {Ward} identity (\ref{gwic}).

Although the approach of  {algebraic renormalization} does not rely on the
choices of renormalization schemes, we choose
the BPHZ renormalization scheme in \cite{Zimmer2} to define the scalar field theory. 
The main reason is that in this scheme insertions can be realized by the normal product
algorithm, see \cite{Low1}, \cite{Zimmer3}, \cite{Zimmer4}.  Then we can directly 
calculate insertions in detail instead of using algebraic constraints. Furthermore,
we can use the  {Zimmermann} identities defined in the BPHZ renormalization 
scheme which relate insertions with different subtraction degrees, see \cite{Zimmer3}. 
As a matter of fact, however, the central results in the paper are independent of the
scheme choices.

The S-matrix is given by amputation of external propagators of the  {Green} function 
in the on-shell limit in the LSZ reduction procedure, which suggests 
that the breaking of the conformal invariance have to be first controlled at the level
of the  {Green} functions. Actually, they are determined by insertion of the trace of 
the energy-momentum tensor which is related to the local \cs equation. With it in hand, 
we can calculate the  {Ward} identities for the conformal transformations to all 
orders in loop expansions. By integrating both sides, the \cs equation, 
see \cite{Callan, Symanzik}, can be obtained and directly related to 
the dilatation transformation. The special conformal transformation of the 
 {Green} function is obtained in a similar manner. Afterwards, the conformal 
transformations of the S-matrix can be calculated by applying the LSZ 
reduction formula.

Furthermore, the LSZ reduction procedure can be used to
construct the charges responsible for the conformal transformations with the 
local  {Ward} identities. For example, the charges for the $\mathit{BRST}$ 
transformations have been obtained in \cite{Kugo1, Kugo2}.
The commutators between the charges and the S-matrix operator 
are to represent the conformal transformations of
the S-matrix in the functional formalism.

Moreover, we will introduce local coupling 
instead of the coupling constant in order to control the conformal breaking 
further. The massive $\phi^4$ model with local coupling can 
be constructed by means of {Poincar{\' e}} invariance and power counting 
renormalizability. The local \cs equation is also calculated then applied to the
calculation of the conformal transformations of the S-matrix.

In addition, we try to obtain the conformal transformations of the
S-matrix in the massless case. In a general sense, it does not
exist due to infra-red divergence. But in \cite{Zimmer1} by
imposing some necessary physical postulates, the S-matrix in the
massless $\phi^4$ field theory can be proved to exist in a
nonperturbative way. Here, we directly assume that it exists and that it
can be obtained by taking the massless limit of the S-matrix
defined in the massive model. Naturally, such a treatment 
is purely formal.


The plan of this paper is the following. In the second section, 
a general procedure to solve our problem is proposed. In the third section, 
we study the conformal transformations of the \sm in the massive $\phif$ defined 
by the BPHZ renormalization procedure. In the fourth section, the conformal 
transformations of the \sm are given by the commutators between the S-matrix operators 
and the corresponding charges. In the fifth section, the conformal transformations 
of the \sm in the massive $\phif$ with local coupling are calculated.
In our conclusion, some remarks suggest that three methods employed
in this paper are also suitable for other models such as non-abelian gauge field
theories or supersymmetrical gauge field theories.
In Appendix A, the proof supporting that the dilatation transformation
of the S-matrix goes without on-shell poles is presented. In Appendix B, 
the problem whether the special conformal transformation
of the S-matrix has on-shell poles or not is discussed up to two-loop order.
In Appendix C, current constructions and  charge constructions 
via the local  {Ward} identities are presented in detail.

\section{The general procedure for calculating the conformal transformations of 
the S-Matrix}

In this section, we propose the general procedure by showing an example of how to 
construct the dilatation transformations of the \sm in the massive $\phif$.

The \sm is constructed from the  {Green} function by the amputation of its external
propagators in the on-shell limit.
With the LSZ reduction procedure, the \sm in the momentum space is given by the 
following expression,
\eq
S_n=\lim_{p_{i}{\in} P}(-ir^{-\half})^n
  \prod_{i=1}^{n}({p_{i}}^2-m^{2})\,G_{n}(p_{1},p_{2}, \cdots,p_{n}),
\en
where
 $\sn$ is the $n$-particle scattering matrix element;
 $\gnp$ is the $n$-particle general  {Green} function, but here we only take the 
 connected part which contains the factor $\delta^4(p_1+p_2+\cdots+p_n)$;
 $r$ is the wavefunction renormalization constant;
$p_{i}$ is the momentum of ith particle,  $m$ is the mass of the scalar particle;
  and $P$ is the set defined by
\eq
  P:=\cp.
\en
The wavefunction renormalization constant $r$ is also defined by
\eq
 {\frac {1}  {r}}:=\partial_{p^2}\widetilde{\Gamma}_{2}{(p,-p)}\mid_{p^2=m^2},
\en
where
 $\widetilde{\Gamma}_{2}{(p,-p)}$ is the two-point $1PI$(one particle irreducible)
  {Green} function.

The dilatation transformation of the \sm can be defined by
\eq
\mathcal{W}^D\,S_n:=\summ\,\sn,
\en
$\mathcal{W}^D$ being the  {Ward} identity operator for the dilatation transformation.
With the definition of the S-matrix, first, we have to control the breaking of
the dilatation invariance of the  {Green} function, namely we have to calculate
\eq
\mathcal{W}^D\,G_n:=\sumx\,\gn.
\en
Second we shall treat the derivative of the following type,
\eq
\label{onshellderivative}
\partial_{p_{i}}\left\{f(p_{1},\cdots,p_{n})\mid_{P}\right\},
\en
since the \sm is obtained by taking the on-shell limit. 
It is observed that there is no direct access to the
derivative (\ref{onshellderivative}), because in the general case
we have
\eq
\partial_{p_{i}}\left\{f(p_{1},\cdots,p_{n})\mid_{P }\right\}\neq
\partial_{p_{i}}f(p_{1},\cdots,p_{n})\mid_{P}.
\en
The derivative 
$p_{i}\partial_{p_{i}}\left\{f(p_{1},\cdots,p_{n})\mid_{P}\right\}$ is
not well-defined since the on-shell condition $p_{i}^2$$=$$m^2$ means that
$p_i^0$, $p_i^1$, $p_i^2$, $p_i^3$ are not independent of each other, but
$\partial_{p_{i}}f(p_{1},\cdots,p_{n})\mid_{P}$ is well-defined and hence
can be used to define the previous one.

Introducing two new functions, 
\eqa
G_{A,n}&:=&\,\gnaa,\\
S_{A,n}&:=&\,\snaa,
\ena
the \sm element $S_n$ is obtained to be found to be
\eq
S_n=\lim_{p_{i}{\in} P}\,\sna,
\en
which implies that 
we can construct $\mathcal{W}^D\,S_n$ with  $\gW^D S_{A,n}\mid_{P}$. 
Hence it is necessary to understand what $ \gW^D S_{A,n}\mid_{P}$ 
does mean in a physical sense. 

Furthermore, for convenience, we introduce the notation $\F^A_n(x;p)$ to denote 
the action of both the  {Fourier} transformation and the amputation, namely
\eq
\F^A_n(x;p):=(ir^{-\half})^n\fur\prod_{i=1}^n(\Box_{x_i}+m^2),
\en
where $(x;p)$ is the abbreviation of $(x_1,\cdots x_n; p_1, \cdots p_n)$.
Then  $S_{A,n}$ is denoted by $\F^A_n(x;p)G_n$. We also introduce
the notation $\F^A_n(x,\check{x}_l;p)$:
\eq
\F^A_n(x,\check{x}_l;p):=(ir^{-\half})^n\fur\prod_{i=1,i\neq l}^n(\Box_{x_i}+m^2).
\en
Similarly,  $\F^A_n(x;-q)$ and $\F^A_n(x,\check{x}_l;-q)$ are given by replacing 
the momentum $p_i$ with $-q_i$ in the corresponding parts.

The general procedure for calculating
the conformal transformations of the \sm can be concluded as follows.
As a starting point, we must have a set of well-defined  {Green} functions. 
Then we carry out the following steps.
\begin{enumerate}
\item  Calculate $\mathcal{W}^D\,\gn$;
\item  Calculate $\mathcal{W}^D\,\gna$;
\item  Calculate $\mathcal{W}^D\,\sna$;
\item  Calculate $\mathcal{W}^D\,\sn$.
\end{enumerate}

In order to check the result, we can compare $\mathcal{W}^D\,\sn$ with known results
in case they exist. If we can control the breaking further, then we have to repeat 
all the above steps. Moreover, there are several reasons for introducing the above procedure. 
First of all, they are set up for solving the puzzle of how to define the
derivation on the on-shell objects. Second, carrying out them step by step will be
helpful to choose suitable formalisms of the conformal transformations as 
realizations of differential operators. Third, like what we will do in the fourth section, 
the conformal transformations of the S-matrix can be also calculated by the commutators 
between charges and the S-matrix operator. However, as will be shown, to define charges 
for the conformal transformations is not an easy task.


\section{Conformal transformations of the \sm \\
   in the massive scalar model}


A well-defined perturbative quantum field theory provides at least exact rules to 
compute renormalized  {Green} functions and derive relations among different 
renormalized  {Green} functions. In this section, the conformal transformations of the 
\sm in the well-defined massive $\phif$ will be calculated. The massive $\phif$ will 
be defined in the BPHZ renormalization scheme where the insertions of the composite 
operators are represented by  {normal products} in \cite{Zimmer2}. 
However,  most results can be also obtained from other regularization schemes or 
renormalization  procedures.

\subsection{The massive $\phif$ via the BPHZ renormalization procedure}

In this subsection, the well-defined massive $\phif$ in the BPHZ renormalization 
scheme is introduced. Namely,  the renormalized action, the renormalization conditions,
the  {Zimmermann's} forest formula, the renormalized  {Green} function, 
the renormalized  {Green} function  with the insertions of the  {normal products}, 
the \QAP and the  {Zimmermann} identities are presented.

The renormalized action $\Gren$ can be regarded as the sum of
the free part $\Gamma_0$ and the interaction part $\Gamma_{\rm int}$,
\eq
\Gren=\Gamma_0+\Gamma_{\rm int}=-z\,\Del{1}-\,a\Del{2}-{\rho}\,\Del{4}.
\en
In the tree approximation, the coefficients $z$, $a$, $\rho$ are specified by  
\eq
z^{(0)}=1, \qquad a^{(0)}=m^2, \qquad  \rho^{(0)}=\lambda,
\en 
where the upper indices denote the power counting of $\hbar$ in this section.
The normal products $\Del{1}$, $\Del{2}$, $\Del{4}$ are given by
\eqa
\Del{1}&=&\delk,\\
\Del{2}&=&\delm,\\
\Del{4}&=&\deli.
\ena
The free part $\Gamma_0$ is given by $-\Del{1}-m^2\Del{2}$
determining the propagator. The renormalized  {Lagrangian} density is chosen 
to be
\eq
\Lren=-\half\,z\,\phi\Box\phi-\,\half\,a\phi^2-{\frac 1 {4!}}\rho\phi^4,
\en
but it can be changed by adding total derivatives.
On the other hand, the renormalized action $\Gren$ is also the sum of the
classical action $\Gamma_{\rm cl}$ and all the possible local counterterms 
$\Gamma_{\rm counter}$, namely
\eq
\Gren=\Gamma_{\rm cl}+\Gamma_{\rm counter}=
-\int (\half\,\phi(\Box+ m^2)\phi+{\frac 1 {4!}}\lambda\phi^4)+\mathcal{O}(\hbar).
\en

In higher orders, the coefficients $z$, $a$, $\rho$ are decided 
by the renormalization conditions
\eqa
\label{normalization}
\widetilde{\Gamma}_{2}{(p,-p)}\mid_{p^2=m^2}&=&0, \\
\partial_{p^2}\widetilde{\Gamma}_{2}{(p,-p)}\mid_{p^2={\mu}^2}&=&1, \\
\widetilde{\Gamma}_{4}{(p_{1}, p_{2}, p_{3},p_{4})}
    \mid_{Q}&=&-\lambda,
\ena
where $m$ is the physical mass, $\mu$ is the normalization mass denoting the 
renormalization scale, $\lambda$ is the physical coupling constant and
$Q$ is the set given by
\eq
Q={\{ p_{i}\mid p_{i}^2={\mu}^2,(p_{i}+p_{j})^2={\frac {4}{ 3}}{\mu}^2; i\neq
j; i,j=1,2,3,4\}}.
\en
 $\widetilde{\Gamma}_{2}{(p,-p)}$ and
 $\widetilde{\Gamma}_{4}{(p_{1}, p_{2}, p_{3}, p_{4})}$ 
are the two-point $1PI$  {Green} function and the
four-point $1PI$  {Green} function respectively. Here
the rule of the  {Fourier} transformation of an ordinary function, 
such as the  {Green} function or 1PI  {Green} function, 
between the momentum space and the coordinate space has been chosen as
{\setlength\arraycolsep{2pt}\eqa
F(p_1, p_2, \cdots, p_{n})
&=& (2\,\pi)^4\,\delta^4\left(\sum_{i=1}^n\,p_i\right)\,
\widetilde{F}(p_1, p_2, \cdots, p_{n}) \nonumber\\
&=&\fur\,F(x_1, x_2, \cdots, x_{n}).
\ena }

The  {Zimmermann} forest formula was given in \cite{Zimmer2}.
It denotes a procedure for obtaining the
renormalized  {Feynman} integrand $R_{\Ga}(p, k)$ from the unrenormalized  {Feynman}
integrand $I_{\Ga}(p, k)$, namely
\eq
R_{\Ga}(p, k)=
\sum_{U\in \F}\,S_{\Gamma}\prod_{\gamma\in U}\,(-t^{\delta_{\gamma}}S_\gamma)\,
\,I_{\Ga}(p, k),
\en
where $U$ is a forest, $\F$ is a set of all possible renormalization
forests, $S_\Gamma$ or $S_\gamma$ are substitution operators, 
$t^{\delta_{\gamma}}$ is the  {Taylor} subtraction operator
cut off by the subtraction degree $\delta_{\gamma}$, the argument
$p$ denotes a set of external momenta and the argument $k$ denotes 
a set of independent internal momenta.

The renormalized  {Green} function in the BPHZ renormalization procedure  
was defined in \cite{Zimmer3, Zimmer4}.
The unrenormalized  {Green} function is given in the Gell-Mann$-$Low formulation
and its renormalized version is directly defined as the finite part under the BPHZ
renormalization procedure,
\eqa
&&\l T\phi(x_1)\phi(x_2)\cdots\phi(x_n)\r \nonumber\\
&&:=\BPHZ~\mbox{finite part ~of~}\nonumber\\
&&
 \l T\phi_0(x_1)\phi_0(x_2)\cdots\phi_0(x_n)e^{{\frac {i} {\hslash}}{\Gamma}^{0}_{\rm int}}\r
 /\l T e^{{\frac {i} {\hslash}}{\Gamma}^{0}_{\rm int}}\r.
\ena
The renormalized  {Green} function with insertions of  {normal products} is given by
 \eqa
 &&\prod_{i}N_{\delta_{i}}\left[Q_{i}(y_i)\right]\cdot\gn  \nonumber\\
&:=&\langle T\prod_{i}N_{\delta_{i}}\left[Q_{i}(y_i)\right]
     \phi(x_1)\phi(x_2)\cdots\phi(x_n)\rangle  \nonumber\\
     &=&\BPHZ~\mbox{finite part ~of~}   \nonumber\\
   && \langle T\prod_{i}N_{\delta_{i}}\left[Q^{0}_{i}(y_i)\right]
   \phi_0(x_1)\phi_0(x_2)\cdots\phi_0(x_n)
   e^{{\frac {i}{\hslash}} {\Gamma}^{0}_{\rm int}}\rangle
   /\l T e^{{\frac {i} {\hslash}}{\Gamma}^{0}_{\rm int}}\r ,
\ena
where the normal products $N_{\delta_{i}}\left[Q^{0}_{i}(y_i)\right]$ denote vertices 
to be treated with the subtraction degree $\delta_{i}$. 
The symbols with the upper index $0$ are defined in free quantum field theory. 
For convenience, the subtraction degrees of normal products will not be explicitly given 
in some cases.

The  {quantum action principle} was derived in the BPHZ
renormalization procedure, see \cite{Low1, Low2, Lam1, Lam2}. 
It means that variations of parameters or fields of the 
 {Green} function can be represented by appropriate local insertions. 
Its differential formalism is given in the following way,
\eqa
\partial_{A}\Gamma&=&\left[\partial_{A}\Gren\right]_4\cdot\Gamma,~~~~~
  A=m, {\mu}, \lambda,\\
{\phi(x){\frac{\delta\Gamma}{\delta\phi(x)}}}&=&
\left[\phi(x){\frac{\delta\Gren}{\delta\phi(x)}}\right]_4\cdot\Gamma,
\ena
where the lower indices of  {normal products} are the subtraction degrees used in the
BPHZ renormalization scheme.

The  {Zimmermann} identities were proved in \cite{Zimmer3, Zimmer4}. 
They relate the subtraction and the over-subtraction in the BPHZ renormalization 
procedure. They are given by 
\eq
N_\delta\left[Q\right]\cdot\Gamma=\sum_{i}u_{i}N_\chi\left[Q_i\right]\cdot\Gamma,
\en
where the sum is over all possible normal products of the over-subtraction degree $\chi$ 
with same quantum numbers, and $\delta$ is the subtraction degree, with 
$\chi$$>$$\delta$. 
The coefficients $u_{i}$ are determined by normalization conditions 
on insertions of composite operators.


\subsection{The local \cs equation}

First of all, we have to control the breaking of the conformal invariance
of the  {Green} function. Choosing one suitable momentum construction
of the local  {Ward} identity operator for the translation transformation as
\eq
\widetilde{\mathbf{w}}_{\mu}^T(x)=
\partial_{\mu}\phi(x){\frac{\delta}{\delta\phi(x)}}-
{\frac {1} {4}}\partial_{\mu}\left(\phi(x){\frac{\delta}{\delta\phi(x)}}\right),
\en
the breaking of the conformal invariance will be determined by
the insertion of the trace of the energy-momentum tensor $T_{\mn}$ given by
\eqa
T_{\mn}=z\partial_{\mu}\phi\partial_{\nu}\phi-\eta_{\mn}({\frac 1 2}z\partial\phi
\partial\phi+{\frac 1 4}z\phi\Box\phi-{\frac 1
4}a\phi^2)-c(\partial_{\mu}\partial_{\nu}-\eta_{\mn}\Box)\phi^2,
\ena
where $T_{\mn}$ satisfies
$\widetilde{\mathbf{w}}_{\mu}^T(x)\Gren=-\partial^{\nu}[T_{\mn}(x)]_4$,
$\eta_{\mn}$ is a metric given in  {Minkowski} space-time
and $c$ is a constant determined by coupling the energy-momentum tensor 
$T_{\mn}$ to a curved background, see \cite{elisabeth1, elisabeth2}. 

The dilatation transformation of the vertex functional $\Gamma$ is defined by
\eqa
\mathcal{W}^D\Gamma[\phi]&:=&\int \dx
(1+x\partial_x)\phi(x)\,{\fdq{\Gamma[\phi]}{\phi(x)}}\nonumber\\
& =&\int\dx [T^{\nu}_{\nu}]_4\cdot\Gamma.
\ena
The special conformal transformation of the vertex functional $\Gamma$
is defined by
\eqa
\alpha\mathcal{W}^K\Gamma[\phi]&:=&\int \dx
\left(\alpha^{\nu}(2x_{\nu}x^{\mu}-\eta^{\mu}_{\nu}x^2)\partial_{\mu}+
2\alpha x\right)\phi(x)\,{\fdq {\Gamma[\phi]} {\phi(x)}}\nonumber\\
& =&\int\dx \,(2\alpha x)\,[T^{\nu}_{\nu}]_4\cdot\Gamma,
\ena
where the symbol $\alpha$ is a constant parameter and $\alpha\mathcal{W}^K$ denotes
the scalar product of $\alpha^\nu\mathcal{W}^K_\nu$.

Hence we have to first carry out the insertion of the trace of the energy-momentum tensor
into the vertex functional $\Gamma$. With the known $\beta_{\lambda}$ function and the
anomalous dimension $\gamma$ used in the \cs equation, the local \cs equation is
obtained by
\eqa
\label{localcs}
&&-[T^{\nu}_{\nu}]_4\cdot\Gamma+
\beta_{\lambda}[\partial_\lambda \Lren]_4\cdot\Gamma
-\half\gamma\phi{\frac {\delta} {\delta \phi}}\Gamma
\nonumber  \\
&&=(z-6c-\half \alpha_m s)[\phi\Box\phi+\partial\phi\partial\phi]_4\cdot\Gamma+
\half \alpha_m [\phi^2]_2\cdot\Gamma,
\ena
where $\alpha_m$ is a parameter to be determined, and we have used the 
 {Zimmermann} identity given by
\eq
\label{zimmeridencons}
\half[\phi^2]_2\cdot\Gamma=\half[\phi^2]_4\cdot\Gamma+
\half s\,[\partial\phi\partial\phi]_4\cdot\Gamma+\half r\,[\phi\Box\phi]_4\cdot\Gamma+
{\frac 1 {4!}} t\,[\phi^4]_4\cdot\Gamma,
\en
the parameters $s$, $r$, $t$ being fixed by the normalization conditions on the
insertions of the related composite operators. 
By defining $\alpha_m$, $\beta_{\lambda}$ and $\gamma$ as the
solutions of the following three equations:
\eq
\label{constraintcondition}
\left\lbrace\begin{array}{lll}
\alpha_m &=&-2a-\beta_{\lambda}\parl a +\gamma a,\\
{\alpha_m}(r-s)&=&-\beta_{\lambda}\parl z+\gamma z,\\
{\alpha_m}t&=&-\beta_{\lambda}\parl\rho+2\gamma\rho,
\end{array}
\right.
\en
we arrive at the conventional form of the  {Callan}--{Symanzik}
equation.

The \cs equation like the ordinary one can be obtained by
integrating the both sides of the equation (\ref{localcs}), \eq \label{cs}
\left(m\partial_m+\beta_{\lambda}\partial_{\lambda}-\half\gamma\N\right)\Gamma=
\alpha_m\,\Del{d}\cdot\Gamma, \en where the symbol
$m\partial_m$ denotes $m\partial_m+\mu\partial_\mu$, the
classical approximation of $\alpha_m$ is given by
$\alpha_m^{(0)}$$=$$-2m^2$ and $\N$ and the $\Del{d}$ are denoted by
\eq
\N=\int\dx\phi(x){\frac{\delta}{\delta\phi(x)}},\qquad \Del{d}=\delmm,
\en
respectively.

Three remarks are given in order. First, there are four differential operators,
\eq
m\partial_m,\,\, \mu\partial_\mu,\,\, \beta_\lambda\partial_\lambda,\,\, \N,
\en
and one identity,  the  {Zimmermann} identity, but we only have
three independent integral insertions. So, we have to obtain two constraint equations.
One is just the $\cs$ equation and the other is the renormalization group equation.
Second, although here the $\cs$ equation is derived in the BPHZ scheme, its formulation
(\ref{cs}) is ordinary, independent of used schemes. Third, taking
the massless limit in a formal sense, we have $\alpha_m$$\to$ $0$ due to
\eq
\label{alpham}
\alpha_m=-2m^2\,{\frac {1} { r\, \Delta_d\cdot\widetilde{\Gamma}_{2}|_{p^2=m^2}} }
\en
as explained in Appendix A.

\subsection{The  {Poincar\'e} transformation of the \sm}

The  {Poincar\'e} transformations consist of translations and 
 {Lorentz} rotations. That the \sm is invariant under  {Poincar\'e} 
transformations
is one fundamental physical requirement in axiomatic quantum field
theory. We have to realize it in our approach.

The \gWI for space-time translations in the generating functional $Z[J]$ 
is defined by
\eq
\mathcal{W}^T_{\mu}\Z[J]:=\int \dx J(x)\,\partial_{\mu}\,{\fdq{\Z[J]}{J(x)}}.
\en

Similarly, the \gWI for  {Lorentz} transformations is defined by
\eq
\mathcal{W}^L_{\mn}\Z[J]:=
\int \dx J(x)\,(x_{\mu}\partial_{\nu}-x_{\nu}\partial_{\mu})\,{\fdq{Z[J]}{J(x)}}.
\en

Since the renormalized action $\Gren$ is invariant under space-time translations and
 {Lorentz} rotations, applying the  {quantum action principle}, we obtain
\eq
\mathcal{W}^T_{\mu}\gn=0,\qquad  \mathcal{W}^L_{\mn}\gn=0.
\en
Due to the commutativity between the  {Klein-Gordon} operator $\Box_x+m^2$
and the differential operators
$\partial_{\mu}$ or \ds{x_{\mu}\partial_{\nu}-x_{\nu}\partial_{\mu}}, we have
\eq
\mathcal{W}^T_{\mu}\gna=0,\qquad  \mathcal{W}^L_{\mn}\gna=0,
\en
which are transformed into the the momentum space formulation by
\eq
\mathcal{W}^T_{\mu}S_{A,n}:=\sum_{l=1}^n p_{l,\mu}S_{A,n}=0,\qquad
\mathcal{W}^L_{\mn}S_{A,n}:=\sum_{l=1}^n
\left(p_{l,\mu}{\frac {\partial} {\partial{p_{l,\nu}}}}-
p_{l,\nu}{\frac {\partial} {\partial{p_{l,\mu}}}}
 \right)S_{A,n}=0.
\en
From theses equations, we can derive the conservation of four-momentum and 
the fact that $S_{A,n}$ is  {Lorentz} invariant. Then we assign to $S_{A,n}$ 
and $S_n$ two new  {Lorentz} invariant functions $S^\prime_{A,n}$ and $S^\prime_n$
in the following way:
\eq
S_{A,n}=\delta^4(\sum_{i=1}^n\,p_i)S^\prime_{A,n}(p_i^2, p_i\cdot p_j, m^2),
\qquad S_n=\delta^4(\sum_{i=1}^n\,p_i)S^\prime_n(p_i\cdot p_j, m^2),
\en
where the second one implies that the \sm is  {Poincar\'e} invariant, namely,
\eq
\mathcal{W}^T_{\mu}S_n=0,\qquad  \mathcal{W}^L_{\mn}S_n=0.
\en

Furthermore, we define $\sum_{i=1}^n\,p_i\partial_{p_i}\,S_n$ by
\eq
\sum_{i=1}^n\,p_i\partial_{p_i}\,S_n:=\sum_{i=1}^n\,p_i\partial_{p_i}\,S_{A,n}|_P-
2m^2 \delta^4(\sum_{i=1}^n\,p_i)\sum_{i=1}^n\,\partial_{p_i^2}\,S^\prime_{A,n}|_P,
\en
$p_i^2$ and $p_i\cdot p_j$, $i\neq j$,
being regarded as independent variables before the on-shell limit.
When we formally take the massless limit in our approach, we find
\eq
\sum_{i=1}^n\,(1+p_i\partial_{p_i})\,S_n=\sum_{i=1}^n\,(1+p_i\partial_{p_i})\,S_{A,n}|_P.
\en

\subsection{The dilatation transformation of the \sm}

In this subsection, the dilatation transformation of the \sm will be treated
with the general procedure proposed above. It is well known that 
the breaking of the dilatational invariance can be characterized by the beta function
$\beta_{\lambda}$ and the anomalous dimension $\gamma$ in the \cs equation. 
We will read off our result in the massless limit and compare it 
with  {Zimmermann's} result in \cite{Zimmer2}.

Define the global  {Ward} identity for the dilatation transformation of 
the generating functional $Z[J]$ by
\eq
\mathcal{W}^D\Z[J]:=\int \dx J(x)\,(1+x\partial_x)\,{\fdq{\Z[J]}{J(x)}},
\en
then the transformation of the general  {Green} function $G_n$ 
under dilatation is
\eq
\mathcal{W}^D\,G_n=\sumx\,\gn.
\en
By the dimensional analysis,
the dilatation transformation of the  {Green} function is represented by
\eq
\mathcal{W}^D\,G_n=m\partial_m\,\gn,
\en
where we denote $\pmu$ by $m\partial_m$ for convenience.

With the \cs equation realized in the  {Green} function obtained by applying
the $\Legendre$ transformation to the equation (\ref{cs}), 
the dilatation transformation of the n-point  {Green} function 
is obtained as
\eq
\label{dilatationgree}
\mathcal{W}^D\,G_n={\frac {i} {\hbar}}\alpha_m \Del{d}\cdot G_{n}-
(\beta_{\lambda}\partial_{\lambda}+\half\,n\gamma) \,G_{n},
\en
where the insertion $\alpha_m \Del{d}\cdot G_{n}$  vanishes in the limit of
large momenta due to the  {Weinberg} asymptotic theorem \cite{wein}
and is thus called ``soft'' breaking of the dilatation invariance.

Calculating $\mathcal{W}^D\,G_{A,n}$, we obtain
\eqa
 \mathcal{W}^D\,G_{A,n}&:=&\sumx{(i r^{-1/2})}^n\prodn \,G_{n}\nonumber\\
    &=&{(i r^{-1/2})}^n
 \prodn \mathcal{W}^D\,G_{n} \nonumber\\
 & &-{(i r^{-1/2})}^n
 \sum_{l=1}^n 2\Box_{x_l}
 \prodl \,G_{n},
\ena
where we used the commutator $\left[1+x\partial_x,\Box_x+m^{2}\right]=-2\Box_x$
implying that the amputation does not commute with the dilatation transformation.

Replacing $\mathcal{W}^D\,G_{n}$ in the above expression with 
(\ref{dilatationgree}), we have
\eqa
\label{Wgna}
\mathcal{W}^D\,G_{A,n}= 
-\beta_{\lambda}\,\partial_{\lambda}\,G_{A,n}-
\half n(\beta_{\lambda}\partial_{\lambda}\ln{r}+\gamma)\,G_{A,n}-2n\,G_{A,n}
   +{\Delta}^\prime_d\cdot G_{A,n},
\ena
where ${\Delta}^\prime_d\cdot G_{A,n}$ is given by
\eqa
&&{\Delta}^\prime_d\cdot G_{A,n}={(i r^{-1/2})}^n\sum_{l=1}^n\, 2m^2\prodl \,G_{n}
\nonumber \\
&& +(i r^{-1/2})^n {\frac {i} {\hbar} }\alpha_m \prodn\Del{d}\cdot G_{n}.
\ena

Then the dilatation transformation of $S_{A,n}$ is given by
\eqa
\mathcal{W}^D\,S_{A,n}&:=&\summ S_{A,n}=-m\partial_m\,S_{A,n} \nonumber\\
&=&\beta_{\lambda}\,\partial_{\lambda}\,S_{A,n}+
\half n(\beta_{\lambda}\partial_{\lambda}\ln{r}+\gamma)S_{A,n}
  -{\Delta}^\prime_d\cdot S_{A,n},
\ena
where ${\Delta}^\prime_d\cdot S_{A,n}$ is the $Fourier$ transformation of
 ${\Delta}^\prime_d\cdot G_{A,n}$, namely
\eq
{\Delta}^\prime_d\cdot S_{A,n}=\fur{\Delta}^\prime_d\cdot G_{A,n}.
\en
Taking the on-shell limit, we obtain
${\Delta}^\prime_d\cdot S_{n}$ by
\eqa
{\Delta}^\prime_d\cdot S_{n}={\Delta}^\prime_d\cdot S_{A,n}\mid_{ P}.
\ena

Two remarks are in order. In the on-shell limit, it seems that
${\Delta}^\prime_d\cdot S_{n}$ is not a well-defined object. We have to prove that
in ${\Delta}^\prime_d\cdot S_{n}$, all on-shell poles like ${\frac {1} {p_i^2-m^2}}$ 
cancel.  In fact, ${\Delta}^\prime_d\cdot S_{n}$ does not include 
contributions from the insertion of  the local integral $\Delta_d$ 
into external propagators of the  {Green} function $G_n$, which makes 
the upper index $\prime$ meaningful,
\eq
{\Delta}^\prime_d\cdot S_{n}= {\frac {i} {\hbar} }\alpha_m
\lim_{p_i\in P}\F^A_n(x;p)\Del{d}^\prime\cdot G_{n},
\en
and hence the amputation of external propagators in
${\Delta}^\prime_d\cdot S_{n}$ can be well-defined. The proof is given in Appendix A.
Here, we generalize the notation ${\Delta}\cdot S_{n}$ to arbitrary insertions
such as a double insertion like $\Delta_1\cdot\Delta_2$,
\eq
\Delta_1\cdot\Delta_2\cdot S_{A,n}:=\F^A_n(x;p)\Delta_1\cdot\Delta_2\cdot G_{n}.
\en
Second, the term $\beta_{\lambda}\partial_{\lambda}\ln{r}+\gamma $ can be represented 
by the  on-shell normalization conditions. Combining the $\cs$ equation for the 
two-point $1PI$  {Green} function with dimensional analysis, we obtain a very useful 
equation,
\eqa
2(1-p^2\partial_{p^2})\widetilde{\Gamma}_{2}{(p,-p)}+(\beta_{\lambda}\partial_{\lambda}-\gamma)
 \widetilde{\Gamma}_{2}{(p,-p)}
 =\alpha_m \Del{d}\cdot\widetilde{\Gamma}_{2}{(p,-p)}.
\ena
By multiplying the derivative $p^2\partial_{p^2}$ on both sides and taking the on-shell
limit, we have
\eqa
  (\beta_{\lambda}\partial_{\lambda}\ln{r}+\gamma)=
  -2\,r\,m^2\partial_{p^2}\partial_{p^2}\widetilde{\Gamma}_{2}{(p,-p)}\mid_{p^2=m^2}
\nonumber\\
-\alpha_m\,r
\partial_{p^2}\Del{d}\cdot\widetilde{\Gamma}_{2}{(p,-p)}\mid_{p^2=m^2}.
\ena

With the following dimensional analysis:
\eqa
&&\left(\sum_{j=1}^n p_{j}^2\partial_{p_j^2}+\pmmu\right)S_{A,n}\mid_{ P}\nonumber \\
&&=\left(\sum_{j=1}^n p_{j}^2\partial_{p_j^2}+\pmmu\right)S_{n},
\ena
which means the on-shell limit does not change the dimension of $S_{A,n}$, we
derive the transformation of the S-matrix under dilatation by
\eqa
\label{dialtatiotransform}
\mathcal{W}^D\,S_n &=&
\gW^D S_{A,n}\mid_{P}-2\left(\parr{m^2} S_n-\parr{m^2} S_{A,n}\mid_{ P}\right)
\nonumber\\
&=&\beta_{\lambda}\,\partial_{\lambda}\,S_{n}+
\half n(\beta_{\lambda}\partial_{\lambda}\ln{r}+\gamma)S_{n}
 -{\Delta}^\prime_d\cdot S_{n}\nonumber\\
&&-2m^2\delta^4(\sum_{i=1}^n\,p_i)\sum_{i=1}^n\,\partial_{p_i^2}
\,S^\prime_{A,n}\mid_{ P}.
\ena 

Four remarks have to be made. First, the dilatation
transformation of the S-matrix seems to be complicated, but the
expression for the off-shell S-matrix element $S_{A,n}$ is simpler.
It is necessary to find what $\gW^D S_{A,n}\mid_{P}$ does mean.
Second, we can take the complete on-shell normalization
conditions, namely choosing the renormalization scale $\mu$ to be the
same as the physical mass scale $m$. Then the residue $r$ is 
the factor 1 and the anomalous dimension $\gamma$ can be written as
\eq \gamma=
-2\,m^2(\partial_{p^2}\partial_{p^2}\widetilde{\Gamma}_{2} +
{\frac {\alpha_m}  {2m^2}}
\partial_{p^2}\Del{d}\cdot\widetilde{\Gamma}_{2})\mid_{p^2=m^2}.
\en
 When we take the normalization condition for the insertion of
the compositor operator $\Del{d}$ as 
$\Del{d}\cdot\widetilde{\Gamma}_{2}\mid_{p^2=m^2}$$=$$1$,  
the parameter $\alpha_m$ is given by $\alpha_m$$=$$-2\,m^2$,
using (\ref{alpham}). Third, when we formally take the 
massless limit, some terms in the above equation
(\ref{dialtatiotransform}) will vanish, \eq \alpha_m\to 0;~~
(\beta_{\lambda}\partial_{\lambda}\ln{r}+\gamma)\to 0;~~
 {\Delta}^\prime_d\cdot S_{n}\to 0;
\en
our result will become the same as  {Zimmermann's},
\eq
\summ S_{n}=\beta_{\lambda}\,\partial_{\lambda}\,S_{n},
\en
where the anomalous dimension $\gamma$ does not show up, 
but may appear in the renormalization group
equation. If we take the massless limit in a formal sense 
under the complete on-shell normalization
condition, the anomalous dimension will vanish, and we will
obtain the   {Callan}--{Symanzik}
equation by
\eq
(\mu\partial_\mu+\beta_\lambda\partial_\lambda)\Gamma=0,
\en
which implies that the anomalous dimension $\gamma$ only has an effect in the
off-shell case \cite{Zimmer1}. Fourth, since the dilatation transformation changes 
the mass of the particle, it cannot be regarded as a type of symmetry. But we can use 
it to relate two theories with different masses in the  {Fock} space, for example,
\eqa
& & S_n(m_2)-S_n(m_1)=-\left(\beta_{\lambda}\,\partial_{\lambda}+
\half n(\beta_{\lambda}\partial_{\lambda}\ln{r}+\gamma)\right)\int_{\ln{m_1}}^{\ln{m_2}}
{\rm d (\ln{m}) }\,\,S_n   \nonumber\\
& &+\int_{\ln{m_1}}^{\ln{m_2}}{\rm d (\ln{m}) }\, {\Delta}^\prime_d\cdot S_{n}
+\delta^4(\sum_{i=1}^n\,p_i)
\,\int_{m_1^2}^{m_2^2}{\,\rm d m^2}\,\, \sum_{i=1}^n \partial_{p_i^2}
S^\prime_{A,n}\mid_{ P}.
\ena

\subsection{The special conformal transformation of the \sm }

The \gWI for the special conformal transformation in the generating functional $Z[J]$
is defined by
\eq
\alpha\mathcal{W}^K\Z[J]:=
\int \dx J(x)\,\left(\alpha^\nu(2x_{\nu}x^{\mu}-\eta^{\mu}_{\nu}x^2)\partial_{\mu}+
2\alpha x\right)\,{\fdq{\Z[J]}{J(x)}}.
\en
With the commutator,
\eq
\left[\sum_{l=1}^n\{(2x_{\nu}x^{\mu}-\eta^{\mu}_{\nu}x^2)\partial_{\mu}+2x_{\nu}\},
\prod_{i=1}^n(\Box_{x_i}+m^2)\right]=-\sum_{l=1}^n 4x_{l,\nu}\Box_{x_l}
\prod_{i=1,i\neq l}^n(\Box_{x_i}+m^2),
\en
we obtain the special conformal transformation of  the ``amputated"  {Green} function
$G_{A,n}$,
\eqa
\label{amputateggreen}
\alpha\mathcal{W}^K G_{A,n}&=&(ir^{-\half})^n\prod_{i=1}^n(\Box_{x_i}+m^2)
\alpha\mathcal{W}^K G_n   \nonumber\\
& &-(ir^{-\half})^n\sum_{l=1}^n (4\alpha x_l)\Box_{x_l}
\prod_{i=1,i\neq l}^n(\Box_{x_i}+m^2)G_n.
\ena

Applying the local \cs equation (\ref{localcs}),
the special conformal transformation of the n-point  {Green} function is calculated
as follows
\eq
\alpha\mathcal{W}^K G_n={\frac i \hbar}\alpha_m\,\alpha\Delta_k\cdot G_n-
{\frac i \hbar}\beta_\lambda
[\partial_\lambda(\alpha\Gren^k)]\cdot G_n-
\half \gamma\sum_{l=1}^n (2\alpha x_l) G_n,
\en
where the insertion $\alpha\Delta_k\cdot G_n$ and $\alpha\Gren^k$ 
denote
\eq
\alpha\Delta_k\cdot G_n=\int \dx(2\alpha x)\half [\phi^2(x)]\cdot G_n,\qquad
\alpha\Gren^k=\int \dx(2\alpha x)[\Lren]_4.
\en
Defining the special conformal transformation on $S_{A,n}$ by 
\eq
\label{specconformal}
\alpha\mathcal{W}^K S_{A,n}:=i\sum_{l=1}^n
\alpha^\nu\delta^{k}_\nu(p_l) S_{A,n}, \en where the differential
operator $\delta^{k}_\nu(p_l)$ is given by
 \eq
\delta^{k}_\nu(p)=p_{\mu}(2{\frac{\partial^2} {\partial
p^{\nu}\partial p_{\mu}}}- 
\eta^{\mu}_{\nu}{\frac {\partial^2} {\partial p^{\zeta}\partial p_{\zeta}}})+ 
2{\frac {\partial} {\partial p^{\nu} } }
\en 
with the argument $p_l$, we obtain its result that
\eq 
\label{specialconformalhh}
\alpha\mathcal{W}^K S_{A,n}=
-{\frac i \hbar}\beta_\lambda \F^A_n(x;p)[\partial_\lambda(\alpha\Gren^k)]\cdot G_n 
-\half \gamma \F^A_n(x;p) \sum_{l=1}^n (2\alpha x_l) G_n 
+\alpha\Delta_k^\prime\cdot S_{A,n}, 
\en 
in which the insertion
$\alpha\Delta_k^\prime\cdot S_{A,n}$ is defined as
\eq
\alpha\Delta_k^\prime\cdot S_{A,n}:= 2m^2\,
\sum_{l=1}^n\,\F^A_n(x,\check{x}_l;p)  (2\alpha x_l)\, G_n +{\frac i
\hbar}\alpha_m\,\F^A_n(x;p) \alpha\Delta_k\cdot G_n. 
\en

With the double insertions, the term ${\frac i\hbar}
\beta_\lambda\F^A_n(x;p)[\partial_\lambda(\alpha\Gren^k)]\cdot G_n$
can  be represented by 
\eq 
{\frac i\hbar}\beta_\lambda(\partial_\lambda+{\frac n 2}\partial_\lambda\ln{r})
([(\alpha\Gren^k)]\cdot S_n)
-{\frac i\hbar}\beta_\lambda[(\alpha\Gren^k)]\cdot[\partial_\lambda
\Gren]\cdot S_n,
\en
which together with (\ref{specialconformalhh}) suggests that 
it is not possible to obtain the combinational term of  
$\beta_\lambda\partial_\lambda ln{r}+\gamma$ in the case of the special conformal 
transformation. Via the complete on-shell normalization condition, the term 
containing $\partial_\lambda\ln{r}$ will vanish and the anomalous dimension $\gamma$ 
will be fixed. Formally taking the massless limit, the term 
$\alpha\Delta_k^\prime\cdot S_{A,n}$ will also become zero. The question 
whether the term 
$\alpha\Delta_k^\prime\cdot S_{A,n}\mid_P$ has on-shell poles or not will 
be answered in Appendix B. Similar to the dilatation transformation of the S-matrix, 
we also define $\alpha\mathcal{W}^K S_{n}$ by taking the on-shell limit of 
$\alpha\mathcal{W}^K S_{A,n}$, namely
\eq
\label{spedeoff}
\alpha\mathcal{W}^K S_{n}:=\alpha\mathcal{W}^K S_{A,n}\mid_P-
4\,i\,m^2\sum_{l=1}^n\alpha\mathcal{P}^k_l S^{\prime}_{A,n}\mid_P,
\en 
where the second term vanishes in the formal massless limit
and the differential operator $\mathcal{P}^k_{l,\nu}$ is given
by
\eq\mathcal{P}^k_{l,\nu}:=\left(\pdq{\delta^4}{p_l^\nu}+
 \delta^4 \sum_{m=1}^n p_{m,\nu}\psdq{(p_m\cdot p_l)}
 \right)\psdq{p_l^2},
\en 
the symbol $\delta^4$ denoting the delta function
$\delta^4(\sum_{i=1}^n\,p_i)$.

Finally, in order to control further the breaking of the conformal transformations 
of the S-matrix, a local coupling $\lambda(x)$ will be introduced 
instead of the constant coupling $\lambda$, since it is observed that 
\eq 
\lim_{\lambda(x)\to
\lambda}{\fdq {} {\lambda(x) }} G_n={\frac i \hbar}
\partial_\lambda[\Lren]_4\cdot G_n.
\en

\section{Conformal transformations of the S-matrix in the functional formalism}

In the above sections, we treated our problem in the ordinary functional space 
instead of in the operator formalism. However, it is possible to recover 
information about the operator formalism in our calculation where the S-matrix 
is defined by using the LSZ reduction procedure. 
We construct the charges responsible for the conformal transformations
with the help of the local  {Ward} identities. Via the commutators 
between the charges and the S-matrix operator, the conformal transformations 
of the S-matrix can be represented in the functional formalism in an effective way. 
In addition, a generating functional of the
``amputated"  {Green} functions is at first given so that all the previous 
calculation can be carried out in terms of functionals.

\subsection{The functional for the ``amputated"  {Green} function}

Define the generating functional for ``amputated"  {Green} functions  by
\eq
Z_A[J]:=Z_A[J,j]|_{j=0}=\Sigma[J,j]Z[j]|_{j=0},
\en
where $\Sigma[J,j]$ is given by
\eq
\Sigma[J,j]=\exp\left
\{-i\hbar\int \dx\,J(x)(\Box_x+m^2){\frac {\delta} {\delta j(x)}}\right\}.
\en

This functional can be used to derive the previous results. As an example, 
the special conformal transformation of the ``amputated"  {Green} function is
calculated. The  {Ward} identity for the special conformal transformation 
is defined by
\eq
\alpha\mathcal{W}^K  Z_A[J]:=\int \dx \alpha^{\nu}(2x_\nu x^\mu-\eta_\nu^\mu x^2)
\widetilde{\mathbf{w}}^T_\mu[J]Z_A[J],
\en
where $\widetilde{\mathbf{w}}^T_\mu[J]$ is given by
\eq
\widetilde{\mathbf{w}}^T_\mu[J](x)=J(x)\partial^x_\mu{\frac {\delta} {\delta J(x)}}
-{\frac 1 4}\partial^x_\mu\left({J(x)\frac {\delta} {\delta J(x)}}\right).
\en

By direct calculation, we find
\eqa
\alpha\mathcal{W}^K Z_A[J]&=&\int \dx J(x)(\Box_x+m^2)\left(\alpha^\nu(2x_\nu
x^\mu-\eta_\nu^\mu\, x^2)\partial_\mu+2\alpha x\right) \nonumber\\
&&\times (-i\hbar){\frac {\delta} {\delta j(x)}}Z_A[J,j]|_{j=0} \nonumber\\
&&-\int \dx\,(4\alpha x) J(x)\Box_x(-i\hbar){\frac {\delta} {\delta j(x)}}Z_A[J,j]|_{j=0}.
\ena

Multiplying by the product of the derivatives
$\prod_{i=1}^n{\frac {\delta}{\delta J(x_i)}}$
then taking $J$$=$$0$, we obtain the result same as in (\ref{amputateggreen}).

\subsection{The functional for the S-matrix in the operator formalism}

The S-matrix operator, the generating functional for the S-matrix element, is given by
\eq
\hat{S}[\hat{\phi}_{\rm in}]=:\hat{\Sigma}[\hat{\phi}_{\rm in},J]:Z[J]|_{J=0},
\en
where the symbol $:$ denotes the normal ordering of operator products,
$\hat{\phi}_{\rm in}$ is a free quantum field operator, and
$\hat{\Sigma}[\hat{\phi}_{\rm in},J]$ is given by
\eq
\hat{\Sigma}[\hat{\phi}_{\rm in},J]=\exp{\hat{X}},
\en
the operator $\hat{X}$ being given by
\eq
\hat{X}=(i\,r^{-\half})\int \dx\,\hat{\phi}_{\rm in}(x)(\Box_x+m^2)
(-i\hbar){\frac {\delta} {\delta J(x)}}.
\en

Here, we expand the field operator $\hat{\phi}_{\rm in}(x)$ in the momentum space by
\eq
\hat{\phi}_{\rm in}(x)=\int\,{\rm d}{\tilde k}\,(a(k)e^{-i\,kx}+a^{\dag}(k)e^{i\,kx}),
\en
where the annihilation operator $a(k)$ and the creation operator $a^{\dag}(k)$ satisfy
the commutator relation
\eq
[a(k),\, a^{\dag}(k^{\prime})]=(2\pi)^3\,2\omega_k\,\delta^3(\vec{k}-\vec{k^\prime}),
\en
and the symbols ${\rm d}{\tilde k}$ and $\omega_k$ denote
\eq
 {\rm d}{\tilde k}={\frac {{\rm d}^3k} {(2\pi)^3\,2\omega_k}},
\qquad \omega_k:=\sqrt{k^2+m^2}.
\en
The state $|k_1, k_2,\cdots k_n \rangle$ is
constructed from the vacuum state $|0\rangle$ by
\eq
|k_1, k_2,\cdots k_n \rangle=a^\dag(k_1)a^\dag(k_2)\cdots a^\dag(k_n)|0\rangle.
\en

In the following, we represent the conformal transformations of the S-matrix by
the commutators between the S-matrix operator $\hat{S}$ and the charge operators.
First, we define the charge $\hat{P}_\mu$ for the translation transformation,
the charge $\hat{M}_{\mn}$ for the  {Lorentz} transformations, the charge $\hat{D}$
for the dilatation transformation and the charge $\hat{K}_\nu$ for the special
conformal transformation.  The conformal transformations of the quantum field 
$\hat{\phi}_{\rm in}(x)$ generated by the charges
are given by
\eqa
{\frac i \hbar}[\hat{P}_{\mu},\,\hat{\phi}_{\rm in}(x)]&=&
\partial_{\mu}\hat{\phi}_{\rm in}(x),\nonumber\\
{\frac i \hbar}[\hat{M}_{\mn},\,\hat{\phi}_{\rm in}(x)]&=&
(x_{\mu}\partial_{\nu}-x_{\nu}\partial_{\mu})\hat{\phi}_{\rm in}(x),\nonumber\\
{\frac i \hbar}[\hat{D},\,\hat{\phi}_{\rm in}(x)]&=&
(1+x\partial_x)\hat{\phi}_{\rm in}(x),
\nonumber\\
{\frac i \hbar}[\hat{K}_{\nu},\,\hat{\phi}_{\rm in}(x)]&=&
((2x_\nu x^{\mu}-\eta^{\mu}_{\nu} x^2)\partial_{\mu}+2x_\nu)
\hat{\phi}_{\rm in}(x),
\ena 
and the relevant details are presented in Appendix C. 

Two remarks have to be made. Even in the free theory, 
it is a very delicate subject to define the
charges responsible for the dilatation transformation and the
special conformal transformation because they will change the mass
and we have to work in different  {Hilbert} spaces. The
reason can be seen from the commutative relations 
between the generators $D$, $K_{\nu}$ and $P^2$ 
at the classical approximation, namely
\eqa
\lbrack D, P^2\rbrack  &=&-2\,P^2, \\
\lbrack K_{\nu}, P^2\rbrack  &=&-4\,x_{\nu}\,P^2.
\ena
They imply that only the massless states are conformal invariant, 
see \cite{Callan2}. But we have treated our problem in terms of the 
 {Green} functions without introducing any charges defined in 
the  {Hilbert} space. Furthermore,
we have assumed that the vacuum state $|0\rangle$
is invariant under the conformal transformations, namely, \eq
\hat{P}_\mu|0\rangle=0,\qquad \hat{M}_{\mn}|0\rangle=0,\qquad
\hat{D}|0\rangle=0,\qquad \hat{K}_{\nu}|0\rangle=0. \en

The commutator between the charge $\hat{P}_\mu$ and the S-matrix operator $\hat{S}$
is given by
\eq
[\hat{P}_\mu,\, \hat{S}]=[\hat{P}_\mu,\,:\hat{\Sigma}:]Z[J]|_{J=0}.
\en
By calculating the commutator between $\mathcal{W}^T$ and $\hat{X}$ and observing
\eq
{\frac i \hbar}[\hat{P}_{\mu},\,\hat{X}]=[\mathcal{W}^T_\mu,\,\hat{X}],
\en
we obtain the result 
\eqa
{\frac i \hbar}[\hat{P}_{\mu},\,\hat{S}]&=&[\mathcal{W}^T_\mu,\,:\hat{\Sigma}:]Z[J]|_{J=0}
\nonumber\\
&=&\mathcal{W}^T_\mu:\hat{\Sigma}:Z[J]|_{J=0}-:\hat{\Sigma}:\mathcal{W}^T_\mu Z[J]|_{J=0}
=0,
\ena
where the potential trouble induced by the normal ordering is avoided since 
the charge $\hat{P}_\mu$ does not mix the creation part and annihilation part of the
asymptotic operator, namely
\eq
{\frac i \hbar}[\hat{P}_{\mu},\,\hat{\phi}^{(+)}_{\rm in}(x)]=
\partial_{\mu}\hat{\phi}^{(+)}_{\rm in}(x),\qquad
{\frac i \hbar}[\hat{P}_{\mu},\,\hat{\phi}^{(-)}_{\rm in}(x)]=
\partial_{\mu}\hat{\phi}^{(-)}_{\rm in}(x).
\en
Similarly, the commutator between the charge $M_{\mn}$ and the S-matrix operator
$\hat{S}$ yields
\eq
 [\hat{M}_{\mn}, \hat{S}]=0.
\en
Hence the quantum field theory we are treating is invariant under  
the  {Poincar\'{e}} transformations.

In the case of the dilatation transformation, the commutator between
$\mathcal{W}^D$ and  $\hat{X}$ is given by
\eq
[\mathcal{W}^D,\hat{X}]={\frac i \hbar}[\hat{D},\hat{X}]+2(ir^{-\half})m^2\int\dx
\hat{\phi}_{\rm in}(x)(-i\hbar){\frac {\delta} {\delta J(x)}}.
\en
The commutator between the charge $\hat{D}$ and the S-matrix operator $\hat{S}$ is
calculated to give
\eqa
{\frac i \hbar}[\hat{D},\hat{S}]&=&{\frac i \hbar}[\hat{D},\,\hat{\Sigma}:]Z[J]|_{J=0}
\nonumber\\
&=&-:\hat{\Sigma}:\mathcal{W}^D Z[J]|_{J=0}\nonumber\\
&&-2m^2(ir^{-\half})\int\dx
(-i\hbar){\frac {\delta} {\delta J(x)}}:\hat{\phi}_{\rm in}(x)
\hat{\Sigma}:Z[J]|_{J=0}.
\ena

In the case of the special conformal transformation, the commutator
between the generator $\alpha\hat{K}$ and the S-matrix operator $\hat{S}$ is calculated by
\eqa &&{\frac i \hbar}[\alpha\hat{K},\hat{S}]=
-:\hat{\Sigma}:\alpha\mathcal{W}^K Z[J]|_{J=0} \nonumber\\
&&-2m^2(ir^{-\half})\int\dx \,(2\alpha x)(-i\hbar)
{\frac {\delta} {\delta J(x)}} :\hat{\phi}_{\rm in}(x)\hat{\Sigma}:Z[J]|_{J=0}.
\ena

\subsection{The conformal transformations of the S-matrix}

As an example, we calculate the dilatation transformation of the S-matrix in the
functional formalism. Replacing $\mathcal{W}^D Z[J]$ by the differential operator
$m\partial_m$  and then using the \cs equation in the generating functional $Z[J]$,
we obtain
\eqa
-:\hat{\Sigma}:\mathcal{W}^D Z[J]|_{J=0}&=&
\beta_\lambda:\hat{\Sigma}:\partial_\lambda Z[J]|_{J=0}
+\half\gamma:\hat{\Sigma}:\N Z[J]|_{J=0} \nonumber\\
&&-{\frac i \hbar}\alpha_m:\hat{\Sigma}:\Delta_d\cdot Z[J]|_{J=0},
\ena
where $\N$ is given by $\int\dx\,J(x){\frac {\delta}{\delta J(x)}}$.
Then the dilatation transformation of the S-matrix operator can be represented by
\eqa
{\frac i \hbar}[\hat{D},\hat{S}]=\beta_\lambda\partial_\lambda \hat{S}
+\half(\beta_\lambda\partial_\lambda\ln{r}+
\gamma):\hat{X}\hat{\Sigma}: Z[J]|_{J=0}
-\Delta_d^{\prime}\cdot \hat{S},
\ena
 $\Delta_d^{\prime}\cdot \hat{S}$ denoting 
\eqa
\Delta_d^{\prime}\cdot \hat{S}&=&
{\frac i \hbar}\alpha_m:\hat{\Sigma}:\Delta_d\cdot Z[J]|_{J=0}\nonumber\\
&&+2m^2(ir^{-\half})\int\dx 
(-i\hbar){\frac {\delta} {\delta J(x)}}
:\hat{\phi}_{\rm in}(x)\hat{\Sigma}:Z[J]|_{J=0}.
\ena
In addition, applying $m\partial_m+\beta_\lambda\partial_\lambda$  on the
S-matrix operator $\hat{S}$, we find
\eq
m\partial_m \hat{S}=-{\frac i \hbar}[\hat{D},\hat{S}],
\en
which is useful for charge constructions.

Furthermore, the consistency of all the results in this subsection 
can be checked with those of
the previous ones in the S-matrix element. Taking the incoming state 
by $|q_1,q_2,\cdots q_{n_1}\rangle$, the outgoing state by 
$\langle p_{n_2},\cdots, p_2,p_1|$, we obtain the  result 
\eqa
&&\langle p_{n_2},\cdots, p_2,p_1|
{\frac i \hbar}[\hat{D},\hat{S}]|q_1,q_2,\cdots q_{n_1}\rangle \nonumber\\
&&=\beta_{\lambda}\,\partial_{\lambda}\,S_{n_1\to n_2}+\half(n_1+n_2)(\beta_{\lambda}
\partial_{\lambda}\ln{r}+\gamma)S_{n_1\to n_2}
-\Delta_d^{\prime}\cdot S_{n_1\to n_2}, 
\ena 
where $S_{n_1\to n_2}$ is the matrix element denoted by $\langle p_{n_2},\cdots,
p_2,p_1|\hat{S}|q_1,q_2,\cdots q_{n_1}\rangle$. The S-matrix element $S_n$
treated before is obtained by taking  $n_1$ as zero and $n_2$ as $n$. Due to the fact 
that the conformal transformations are linear, the dilatation transformation of the
field operator $\phi_{\rm in}(x)$ can be decomposed into two 
independent parts, namely
\eq
{\frac i \hbar}[\hat{D},\,\hat{\phi}^{(+)}_{\rm in}(x)]=
(1+x\partial_x)\hat{\phi}^{(+)}_{\rm in}(x),\qquad {\frac i
\hbar}[\hat{D},\,\hat{\phi}^{(-)}_{\rm in}(x)]=
(1+x\partial_x)\hat{\phi}^{(-)}_{\rm in}(x). 
\en
With these in hand, we obtain the matrix element realization of the commutator 
${\frac i \hbar}[\hat{D},\hat{S}]$ by \eqa &&\langle p_{n_2},\cdots, p_2,p_1|
{\frac i \hbar}[\hat{D},\hat{S}]|q_1,q_2,\cdots q_{n_1}\rangle \nonumber\\
&&=\left(\sum_{l=1}^{n_1}(1+q_l\partial_{q_l})+\sum_{l=1}^{n_2}(1+p_l\partial_{p_l})
  \right)S_{A,n_1\to n_2}|_P,
\ena
which gives a meaning to our calculation of the derivatives of 
the following type:
\eq
\partial_{p_{i}}\left\{f(p_{1},\cdots,p_{n})\right\}\mid_{P}.
\en
 Hence we declare
$\mathcal{W}^D\,S_{A,n}\mid_P$ to be the matrix element of the commutator
between $\hat{D}$ and $\hat{S}$.

In the case of the special conformal transformation, we obtain
the result in the operator formalism 
\eqa {\frac i \hbar}[\alpha\hat{K},\hat{S}]&=& {\frac i\hbar}
\beta_\lambda:\hat{\Sigma}:\partial_\lambda(\alpha\Grenk)\cdot Z[J]\mid_{J=0}
\nonumber\\&+& \half\gamma:\hat{\Sigma}:\alpha\N^k
Z[J]|_{J=0} -\alpha\Delta_k^{\prime}\cdot \hat{S}, 
\ena where $\alpha\N^k$ is given by
\eq
\alpha\N^k=\int\,\dx\,(2\alpha x)J(x){\frac \delta {\delta J(x)}},
\en
and $\alpha\Delta_k^{\prime}\cdot \hat{S}$ is defined by
\eqa
&&\alpha\Delta_k^{\prime}\cdot \hat{S}:={\frac i\hbar}
\alpha_m:\hat{\Sigma}:\alpha\Delta_k\cdot Z[J]|_{J=0} \nonumber\\
&&+2m^2(ir^{-\half})\int\dx (2\alpha\,x) :\hat{\phi}_{\rm
in}(x)\hat{\Sigma}:(-i\hbar){\frac{\delta}{\delta J(x)}}Z[J]|_{J=0}. 
\ena
In addition, the above result can be also realized
in the S-matrix element,
\eq
\langle p_{n_2},\cdots, p_2,p_1|{\frac i \hbar}
[\alpha\hat{K},\hat{S}]|q_1,q_2,\cdots q_{n_1}\rangle \nonumber\\
=i\left(\sum_{l=1}^{n_1}\alpha\delta^k(q_l)-\sum_{l=1}^{n_2}\alpha\delta^k(p_l)\right)
S_{A,n_1\to n_2}|_P. \en


\section{Conformal transformations of the S-matrix with local coupling}

In this section, we will treat the conformal transformations of the S-matrix in
the $\phi^4$ model with an external field: the case of local coupling. It was originally 
used to study the renormalizability like in \cite{epstein}. In such a case,
the breaking of the conformal invariance can be controlled  better in principle
than with constant coupling, because the insertion of the trace of the 
energy-momentum can be represented by the action of differential operators. 
It is also helpful for 
constructing charges and carrying out consistency conditions to all orders
\cite{elisabeth2}.

In the following, a well-defined massive $\phi^4$ model with local 
coupling in the BPHZ renormalization procedure is first introduced. Then the local \cs 
equation is calculated and used to derive both the dilatation transformation 
and the special 
conformal transformation of the S-matrix. All results of this section in the   
constant coupling limit are required to return to those of the above sections.

\subsection{The massive $\phi^4$ model with local coupling}

The renormalized action $\Gamma_{{\rm ren}, \lambda}$ is constructed 
by requiring that it is   {Poincar\'e} invariant and satisfies 
dimensional constraints  of the power-counting renormalizability. 
First, we will list all possible independent  {Poincar\'e} invariant 
local basis as monomials of $\lambda(x)$ and $\phi(x)$ with 
dimension four:
{\setlength\arraycolsep{2pt} \eqa
I^{(n)}_m&=&\lambda^n\phi^2,\nonumber \\
I^{(n)}_l&=&\lambda^n\phi\Box\phi,\nonumber \\
I^{(n)}_4&=&\lambda^n\phi^4,\nonumber \\
I^{(n)}_1&=&\lambda^{n-1}\partial_{\mu}\lambda\partial^{\mu}\phi^2,\nonumber \\
I^{(n)}_\lambda&=&\lambda^{n-2}\partial_{\mu}\lambda\partial^{\mu}\lambda\phi^2,\nonumber \\
I^{(n)}_k&=&\half\Box(\lambda^n\phi^2),\nonumber \\
I^{(n)}_2&=&{\frac 1 n}\partial_{\mu}(\partial^{\mu}\lambda^n\phi^2). 
\ena} 

Then the renormalized action is an integral over the space-time variables 
of all possible linear combinations in the  whole above local basis, 
\eq 
\Gamma_{{\rm ren}, \lambda}=\sum_{n=0}^{\infty}\int\,
[-\half z^{(n)}I^{(n)}_l-\half {a^{(n)}} I^{(n)}_m-{\frac 1 {4!}}
\rho^{(n)}I^{(n+1)}_4+\tilde{z}^{(n)}I^{(n)}_1+z^{(n)}_\lambda
I^{(n)}_\lambda], 
\en 
where all coefficients can be determined and the upper indices  
denote the power counting of local coupling in this section. 
In the classical approximation,
we desire that the $\phi^4$ model with local coupling
 returns to the original one, which means 
 \eq
z^{(0)}=1, \qquad a^{(0)}=m^2,\qquad \rho^{(0)}=1, \qquad
\tilde{z}^{(0)}=0, \qquad z^{(0)}_\lambda=0.
 \en
In higher orders, the coefficients $\zn$, $\an$, $\rn$, $n\geq 1$, can be fixed 
by the renormalization conditions similar to (\ref{normalization}),
\eqa
\label{normalizationlocal}
\lim_{\lambda(x)\to\lambda}\widetilde{\Gamma}_{2}{(p,-p)}\mid_{p^2=m^2}&=&0, \\
\lim_{\lambda(x)\to\lambda}\partial_{p^2}\widetilde{\Gamma}_{2}{(p,-p)}\mid_{p^2={\mu}^2}&=&1, \\
\lim_{\lambda(x)\to\lambda}\widetilde{\Gamma}_{4}{(p_{1}, p_{2}, p_{3},p_{4})}
    \mid_{Q}&=&-\lambda,
\ena
where the symbol $\lambda$ denotes the constant coupling. The coefficients $\tzn$,
$\lzn$, $n\geq 1$, can be decided by suitable renormalization conditions
which are not given here since they are not used. In addition, we arrange that 
the relation between the counting number of loops (the power counting of $\hbar$) and 
the counting number of local coupling $\lambda(x)$ is the same as in the case of 
constant coupling, which means that $z^{(1)}_\lambda$$=$$0$.

For the perturbative calculation in higher orders, we still take the BPHZ
renormalization procedure to define the finite  {Green} function and apply the
 {normal product} algorithm to define insertion of composite operators, 
since local coupling is introduced as the external field and this only changes 
the assignment of external momenta. The \QAP with local coupling is given in
its differential formalism,
\eqa
\partial_{A}\Gamma&=&\left[\partial_{A}\Gamma_{{\rm ren}, \lambda}\right]_4\cdot\Gamma,~~~~~
  A=m, {\mu}, \\
 {\frac {\delta\Gamma} {\delta\lambda(x)}}&=&
\left[{\frac{\delta\Gamma_{{\rm ren}, \lambda}}{\delta\lambda(x)}}\right]_4\cdot\Gamma, \qquad 
{\phi(x){\frac{\delta\Gamma}{\delta\phi(x)}}}=
\left[\phi(x){\frac{\delta\Gamma_{{\rm ren},
\lambda}}{\delta\phi(x)}}\right]_4\cdot\Gamma.
\ena
The  {Zimmermann} identity with local coupling is still constructed by expanding 
the insertion of normal product with the lower subtraction degree in a linear combination 
of all possible independent insertions with the same higher subtraction degree. For example,
{\setlength\arraycolsep{2pt} \eqa
\label{zimmerlocal}
&& \half[\phi^2]_2\cdot\Gamma=\half[\phi^2]_4\cdot\Gamma+
\sum_{n=0}^{\infty}[\half u^{(n)}_l\Iln+{\frac 1 {4!}}u^{(n)}_4 I^{(n+1)}_4]\cdot\Gamma  
\nonumber\\
 &&+\sum_{n=0}^{\infty}[u^{(n)}_1\Ion+u^{(n)}_\lambda\Idn+
v^{(n)}_2\Itn+v^{(n)}_k\Ikn]_4\cdot\Gamma,
\ena
which will return to the  {Zimmermann} identity (\ref{zimmeridencons}) in 
the constant coupling limit.

\subsection{The local \cs equation}

Define the energy momentum tensor $T_{\mn}$ by the \lWI for space-time
translations, namely
\eq
\widetilde{\mathbf{w}}_{\mu}^{T}\Gamma[\phi,\lambda]
=:-\partial^{\nu}[T_{\mn}]_4\cdot\Gamma[\phi,\lambda],
\en
where the contact term $\widetilde{\mathbf{w}}_{\mu}^{T}$ is defined by
\eq
\widetilde{\mathbf{w}}_{\mu}^{T}[\phi,\lambda]:=\partial_{\mu}\phi{\frac {\delta} 
{\delta \phi}}-
{\frac 1 4}\partial_{\mu}(\phi{\frac {\delta} {\delta \phi}})+\partial_{\mu}\lambda
{\frac {\delta} {\delta \lambda}}.
\en

With a local coupling, the breaking of the conformal invariance is still controlled by
the insertion of the trace of the energy-momentum tensor which is calculated to be
\eq
T^{\nu}_{\nu}=\sum_{n=0}^{\infty}\an\Imn-
\sum_{n=0}^{\infty}(\zn-6c^{(n)})\Ikn-\sum_{n=0}^{\infty}\tzn\Itn,
\en
where $c^{(n)}$ denote contributions from total derivatives and can be determined 
by the introduction of a curved background like in \cite{elisabeth1, elisabeth2}.
The local \cs equation reads
\eq
[T^{\nu}_\nu](x)\cdot\Gamma-\sum_{k=0}^{\infty}\beta^{(k)}_{\lambda}\lambda^{k+1}(x)
{\frac {\delta} {\delta \lambda(x)}}\Gamma+\half\,\sum_{k=0}^{\infty}\gamma^{(k)}\lambda^{k}(x)
\phi(x){\frac {\delta} {\delta \phi(x)}}\Gamma
=-\mathbf{\Delta}_\lambda(x)\cdot\Gamma,\en
where the insertion $\mathbf{\Delta}_\lambda(x)\cdot\Gamma$ is given by
\eq
\mathbf{\Delta}_\lambda\cdot\Gamma=\sum_{n=0}^{\infty}\half\alpha^{(n)}_m[\Imn]_2\cdot\Gamma
+\sum_{n=0}^{\infty}[A^{(n)}_1\Ion+A^{(n)}_\lambda\Idn+
A^{(n)}_2\Itn+ A^{(n)}_k\Ikn]_4\cdot\Gamma.
\en
We have used the  {Zimmermann} identity with local coupling (\ref{zimmerlocal})
and defined parameters $\alpha_m^{(n)}$, $\beta_\lambda^{(n)}$ and
$\gamma^{(n)}$ as the  solutions of the following three equations,
\eq\left\lbrace\begin{array}{lll}
\alpha^{(n)}_m&=&-2a^{(n)}+\sum_{k=0}^{n}a^{(n-k)}\gamma^{(k)}-\sum_{k=0}^{n}(n-k)
a^{(n-k)}\beta_{\lambda}^{(k)}, \\
\alpha^{(k)}_m u^{(n-k)}_l&=&-\beta^{(k)}_\lambda z^{(n-k)}(n-k)+\gamma^{(k)}z^{(n-k)},
\\
\alpha^{(k)}_m u^{(n-k)}_4&=&-\beta^{(k)}_\lambda(n-k+1)\rho^{(n-k)}+2\rho^{(n-k)}\gamma^{(k)},
\end{array}
\right.
\en
which are consistent with 
(\ref{constraintcondition}) in the constant coupling limit.
Hence the coefficients $A^{(n)}_1$,  $A^{(n)}_\lambda$, $A^{(n)}_2$, $A^{(n)}_k$ are
specified by
{\setlength\arraycolsep{2pt} \eqa
A^{(n)}_1&=&\sum_{k=0}^n(n\beta^{(k)}_\lambda\tilde{z}^{(n-k)}-
\gamma^{(k)}\tilde{z}^{(n-k)})-\sum_{k=0}^{n}\alpha^{(k)}_m
(u^{(n-k)}_1-\half k v^{(n-k)}_k), \nonumber\\
A^{(n)}_\lambda&=&\sum_{k=0}^n[(n+k)\beta^{(k)}_\lambda z^{(n-k)}_\lambda
-\gamma^{(k)}({\frac 1 4}(n-k)k{z}^{(n-k)}+k\tilde{z}^{(n-k)}+z^{(n-k)}_\lambda)]
\nonumber\\
&&-\sum_{k=0}^n\alpha^{(k)}_m[u^{(n-k)}_\lambda-k v^{(n-k)}_2-\half (n-k)k v_2],
 \nonumber\\
A^{(n)}_2&=&\sum_{k=0}^n[\beta^{(k)}_\lambda(n\tilde{z}^{(n-k)}-2 z^{(n-k)}_\lambda)
+\half \gamma^{(k)}(2\tilde{z}^{(n-k)}+\half(n-k)z^{(n-k)})+2\tilde{z}^{(n)}\delta_{n,k}]
\nonumber\\
&&-\sum_{k=0}^n\alpha^{(k)}_m(v^{(n-k)}_2-\half k v^{(n-k)}_k),\nonumber\\
A^{(n)}_k&=&\sum_{k=0}^n[-2\beta^{(k)}_\lambda\tilde{z}^{(n-k)}+
(\zn-6c^{(n)})\delta_{n,k}]-\sum_{k=0}^n\alpha^{(k)}_m v^{(n-k)}_k.
\ena}

In the above formulae, we have used the following results on the beta 
function $\beta_\lambda$ and the anomalous dimension $\gamma$ from a 
perturbative calculation,
\eq
\beta^{(0)}_\lambda=0,\qquad\beta^{(1)}_\lambda=\mathcal{O}(\hbar), \qquad
\gamma^{(0)}=\gamma^{(1)}=0,\qquad \gamma^{(2)}=\mathcal{O}(\hbar^2).
\en
Furthermore, we obtain the \cs equation with local coupling by
\eq
\label{cselocal}
m\partial_m \Gamma_n+\sum_{k=0}^{\infty}\beta_\lambda^{(k)}\int \dx \lambda^{k+1}(x)
{\frac {\delta} {\delta \lambda(x)}}\Gamma_n-\half\sum_{k=0}^{\infty}\gamma^{(k)}
\sum_{l=1}^n\lambda^k(x_l)\Gamma_n=
\mathbf{\Delta}_{d\lambda}\cdot\Gamma_n,
\en
where the normal product $\mathbf{\Delta}_{d\lambda}$ is given by 
$\int {\rm d}^4 x \,\,\Delta_{\lambda}(x)$. In the constant coupling limit, 
it will return to the ordinary \cs equation with the following expansions of 
$\beta_\lambda$ and $\gamma$ in the coupling constant $\lambda$,
\eq
\beta_\lambda=\sum_{k=0}^{\infty}\beta_\lambda^{(k)}\lambda^{k+1}, 
\qquad \gamma=\sum_{k=0}^{\infty}\gamma^{(k)}\lambda^{k}.
\en

In the following, we start to study the conformal transformations of the S-matrix.
Similar to the situation with constant coupling, we obtain the
 {Poincar\'e} transformations of the S-matrix given by 
\eqa
\mathcal{W}^T_\mu S_n&:=&\sum_{l=1}^n p_{l,\mu} S_n=\int \dx \partial_\mu\lambda(x)
{\frac {\delta}{\delta \lambda(x)}}S_n, \\
\mathcal{W}^L_{\mn} S_n&:=&\sum_{l=1}^n
\left(p_{l,\mu}{\frac {\partial} {\partial{p_{l,\nu}}}}-
p_{l,\nu}{\frac {\partial} {\partial{p_{l,\mu}}}}\right)S_{n} \nonumber\\
&=&\int \dx\,
(x_\mu\partial_\nu\lambda(x)-x_\nu\partial_\mu\lambda(x))
{\frac {\delta}{\delta \lambda(x)}}S_n.
\ena

\subsection{The dilatation transformation of the S-matrix}

Define the  {Ward} identity for the dilatation transformation with local
coupling by
\eqa
\mathcal{W}^D Z[J,\lambda]:&=&\int\dx\left(J(x)(1+x^\mu\partial_\mu)
{\frac {\delta} {\delta J(x)}}-x^\mu\partial_\mu\lambda(x)
{\frac {\delta} {\delta \lambda(x)}}\right)Z[J,\lambda]   \nonumber\\
&=& m\partial_m Z[J,\lambda].
\ena

The dilatation transformation of the ``amputated" S-matrix element $S_{A,n}$
is given by
\eqa
\mathcal{W}^D S_{A,n}&:=&\sum_{i=1}^{n}(1+p_i\partial_{p_i})S_{A,n}  \nonumber\\
&&=-\F^A_n(x;p)\mathcal{W}^D G_n -2m^2\sum_{l=1}^n\F^A_n(x,\check{x}_l;p) G_n,
\ena
where $\mathcal{W}^D G_n$ is calculated by $\sum_{l=1}^n(1+x_l\partial_{x_l})G_n$.

With the \cs equation (\ref{cselocal}), we obtain the result 
{\setlength\arraycolsep{2pt} \eqa
\label{dls}
\mathcal{W}^D S_{A,n}&=&\sum_{k=0}^{\infty}\beta^{(k)}_\lambda\int \dx \lambda^{k+1}(x)
{\frac {\delta} {\delta\lambda(x)}}S_{A,n} \nonumber\\
&&+\half \mathcal{B}^d_n S_{A,n}
-\mathcal{H}^d_n(\gamma, {\frac {\delta} {\delta \lambda(x)}})G_n
-\mathbf{\Delta}_{d\lambda}^\prime\cdot S_{A,n},
\ena}
$\mathcal{B}^d_n$ being given by
\eq
\mathcal{B}^d_n=n\sum_{k=0}^{\infty}\beta_{\lambda}^{(k)}\int\dx \lambda^{k+1}(x)
{\frac {\delta}{\delta \lambda(x)}}\ln{r}+
\sum_{l=1}^n\sum_{k=0}^{\infty}\gamma^{(k)}\lambda^{k}(x_l),
\en
and the parameter $r$ being the wavefunction renormalization constant in the coupling
constant limit;
$\mathcal{H}^d_n(\gamma, {\frac {\delta} {\delta \lambda}})$ is given by
\eqa
&&\mathcal{H}^d_n(\gamma, {\frac {\delta} {\delta \lambda}})
=\F^A_n(x;p)\int \dx\,\, x^{\mu}\partial_{\mu}\lambda(x){\frac {\delta} {\delta \lambda(x)}}\nonumber\\
&&-\sum_{l=1}^n \F^A_n(x,\check{x}_l;p)\sum_{k=0}^{\infty}\gamma^{(k)}
\left(\half\Box_{x_l}\lambda^k(x_l)+{\frac {\partial\lambda^k(x_l)} {\partial{x_l^{\mu}}}}
{\frac {\partial} {\partial{x_{l,\mu}}}}\right); 
\ena
and $\mathbf{\Delta}_{d\lambda}^\prime\cdot S_{A,n}$ is given by
\eq
\mathbf{\Delta}_{d\lambda}^\prime\cdot S_{A,n}=
2m^2 \sum_{l=1}^n\F^A_n(x,\check{x}_l;p) G_n+
{\frac i \hbar}\F^A_n(x;p)\mathbf{\Delta}_{d\lambda}\cdot G_n.
\en

In the constant coupling limit, the expression (\ref{dialtatiotransform})
can be obtained via the above equation (\ref{dls}).  The improved \cs operator 
$\hat{\mathcal{C}}$ is defined by
\eq
\hat{\mathcal{C}}:=m\,\partial_m+\sum_{k=0}^{\infty}\beta_\lambda^{(k)}\int\lambda^{k+1}
{\frac {\delta} {\delta \lambda  }}-\half\sum_{k=0}^{\infty}\gamma^{(k)}
\int\lambda^{k}\phi{\frac {\delta} {\delta \phi}},
\en
which is applied to $S_{A,n}$ to obtain
\eqa
\hat{\mathcal{C}}\,S_{A,n}&=&\mathbf{\Delta}_{d\lambda}^\prime\cdot S_{A,n}
\,+\,\mathcal{H}^d_n(\gamma, {\frac {\delta} {\delta \lambda}})\,G_n
\nonumber\\
&&-\half\,n\sum_{k=0}^{\infty}\beta_{\lambda}^{(k)}\int\dx \lambda^{k+1}(x)
{\frac {\delta \ln{r}}{\delta \lambda(x)}}\,S_{A,n}.
\ena

\subsection{The special conformal transformation of the S-matrix}

Define the  {Ward} identity for the special conformal transformation by
\eqa
\alpha\mathcal{W}^{K}Z[J,\lambda]&:=&\int\dx\,
\alpha^\nu(2x_\nu x^\mu-\eta^\mu_\nu x^2)
(J(x)\partial_\mu{\frac {\delta} {\delta J(x)}}-
\partial_\mu\lambda(x){\frac {\delta} {\delta \lambda(x)}})Z\nonumber\\
&&+\int\dx\,(2\alpha x) J(x){\frac {\delta} {\delta J(x)}} Z
\ena

With the definition (\ref{specconformal}) of the special conformal
transformation of the
 ``amputated" S-matrix element $S_{A,n}$, we obtain
{\setlength\arraycolsep{2pt} \eqa
&&\alpha\mathcal{W}^{K} S_{A,n}
=-\sum_{k=0}^{\infty}\beta^{(k)}_\lambda\int \dx\, (2\alpha x)\lambda^{k+1}(x)
{\frac {\delta} {\delta\lambda(x)}}S_{A,n} \nonumber\\
&&-\half \alpha\mathcal{B}^k_n S_{A,n}
+\alpha\mathcal{H}^k_n (\gamma, {\frac {\delta} {\delta \lambda}})G_n
+\alpha\mathbf{\Delta}^{\prime}_{k\lambda}\cdot S_{A,n},
\ena}
where $\alpha\mathcal{B}^{k}_n$ is given by
\eq
\alpha\mathcal{B}^k_n=
n\sum_{k=0}^{\infty}\beta_{\lambda}^{(k)}\int\dx (2\alpha x)\lambda^{k+1}(x)
{\frac {\delta}{\delta \lambda(x)}}\ln{r}+\sum_{l=1}^n 
\sum_{k=0}^{\infty} \gamma^{(k)}\,(2\alpha x_l)\lambda^{k}(x_l);
\en
$\alpha\mathcal{H}^k_n(\gamma, {\frac {\delta} {\delta \lambda}})$ is given by
\eqa
&&\alpha\mathcal{H}^k_n(\gamma, {\frac {\delta} {\delta \lambda}})
=\F^A_n(x;p)\int \dx \alpha^\nu
(2x_\nu x^{\mu}-\eta^{\mu}_{\nu} x^2)\partial_{\mu}\lambda(x)
{\frac {\delta} {\delta \lambda(x)}}\nonumber\\
&&-\sum_{l=1}^n \F^A_n(x,\check{x}_l;p)\sum_{k=0}^{\infty}\gamma^{(k)}
\left(\Box_{x_l}(\alpha x_l\lambda^k(x_l))+
{\frac {\partial (2\alpha x_l\lambda^k(x_l))} {\partial{x_l^{\mu}}}}
{\frac {\partial} {\partial{x_{l,\mu}}}} \right);
\ena
and $\alpha\mathbf{\Delta}^{\prime}_{k\lambda}\cdot S_{A,n}$ is given by
\eqa
\alpha\mathbf{\Delta}^{\prime}_{k\lambda}\cdot S_{A,n}=
2m^2\sum_{l=1}^n \F^A_n(x,\check{x}_l;p) (2\alpha x_l) G_n +
{\frac i \hbar}\F^A_n(x;p)\alpha\mathbf{\Delta}_{k\lambda}\cdot G_n,
\ena
with $\alpha\mathbf{\Delta}_{k\lambda}$ given by 
\eq
\alpha\mathbf{\Delta}_{k\lambda}=\int \dx\,(2\alpha x) \mathbf{\Delta}_{\lambda}(x).
\en

The improved  {Ward} identity operator $\alpha\hat{\mathcal{W}}^K$
of the special conformal transformation is defined by
\eq
\alpha\hat{\mathcal{W}}^K:=\alpha\mathcal{W}^K+
\sum_{k=0}^{\infty}\beta_\lambda^{(k)}\int\,(2\alpha x)\lambda^{k+1}
{\frac {\delta} {\delta \lambda  }}-\half\sum_{k=0}^{\infty}\gamma^{(k)}
\int\,(2\alpha x)\lambda^{k}\phi{\frac {\delta} {\delta \phi}}.
\en
Applying it to $S_{A,n}$, we find 
\eqa
\alpha\hat{\mathcal{W}}^K\, S_{A,n}&=&
\alpha\mathbf{\Delta}^{\prime}_{k\lambda}\cdot S_{A,n} 
+\alpha\mathcal{H}^k_n(\gamma, {\frac {\delta} {\delta \lambda}})\,G_n
\nonumber\\
&&-\half\, n\sum_{k=0}^{\infty}\beta_{\lambda}^{(k)}\int\dx (2\alpha x)\lambda^{k+1}(x)
{\frac {\delta\ln{r} }{\delta \lambda(x)}}\,S_{A,n}.
\ena



\section{Concluding remarks}

In this paper, we followed three approaches to investigate the behaviour 
of the S-matrix under the conformal transformations. 
First, one way of studying the conformal transformations of the S-matrix 
by means of the LSZ reduction procedure is proposed. 
We derive the  {Ward} identities for 
the conformal transformations of the  {Green} functions with the local \cs equation, 
then obtain the conformal transformations of the off-shell S-matrix by calculating 
the commutators 
between  {Ward} identity operators and the  {Klein-Gordon} operator $\Box_x+m^2$, 
and thus
represent the conformal transformation of the S-matrix in terms of the off-shell 
S-matrix 
in the on-shell limit. Second, with charge constructions, we calculate 
the conformal transformations of 
the S-matrix in the functional formalism and realize physical meanings 
of the conformal transformations
of the off-shell S-matrix in the on-shell limit. Third, we also calculate 
the conformal transformations 
of the S-matrix in the case of a local coupling. 
As was shown up, three different types of results are consistent with each other. 

For the dilatation transformation, we obtain the simple result
\eqa
\summ S_{n}&=&\beta_{\lambda}\,\partial_{\lambda}\,S_{n}+
\half n(\beta_{\lambda}\partial_{\lambda}\ln{r}+\gamma)S_{n}
  -{\Delta}^\prime_d\cdot S_{n}\nonumber\\
 & & -2m^2\delta^4(\sum_{i=1}^n\,p_i)\sum_{i=1}^n\,\partial_{p_i^2}\,S^\prime_{A,n}\mid_{ P},
\ena
which yields in the  massless limit
\eq
\summ S_{n}=\beta_{\lambda}\,\partial_{\lambda}\,S_{n}.
\en 
But in the case of the special conformal transformation, the result seems 
to be complicated, which suggests that we have to treat other models such as
supersymmetrical field theories. 

In addition, a proof that the dilatation transformation of the S-matrix has no 
on-shell poles is given. It is independent of the chosen regularization scheme
and renormalization procedure. It is based on  the skeleton expansion, the 
 {Callan}--{Symanzik} equation and the on-shell renormalization conditions.
Furthermore, the discussion whether the special conformal transformation of 
the S-matrix has on-shell poles or not is given in detail. First, 
the problem is simplified by considering the skeleton expansion and 
using conservation of energy-momentum. Then the perturbative calculation 
is carried out up to two-loop.

Some remarks are in order. Firstly, in the framework of the algebraic
renormalization procedure, consistency conditions among the  {Ward} 
identity operators, see \cite{elisabeth2}, can be used to evaluate coefficients in 
the \cs equation with a local coupling. For example, with the help of
the commutativity 
between the \cs operator ${\hat{\mathcal{C}}}$  and the improved  {Ward} identity 
operator $\hat{\mathcal{W}}^K$ of the special conformal transformation, 
the coefficient $A^{(n)}_1$  can be proved to vanish. 

Secondly, the external field $q(x)$ 
can be introduced to control the soft breaking,
\eq
\Delta_d\cdot\Gamma=\int d^4x {\fdq{\Gamma}{q(x)}}\mid_{q(x)=0}.
\en
With two external fields $\lambda(x)$ and $q(x)$, the insertion of the trace of the 
energy-momentum tensor can be completely represented by the action of differential
operators, namely
\eq
  [T^\nu_\nu]_4\cdot\Gamma=\lim_{\lambda(x)\to\lambda}
\left(\beta_\lambda{\frac {\delta}{\delta \lambda(x)}}-\half \gamma\phi(x)
{\frac {\delta}{\delta \phi(x)}}+\half \alpha_m{\frac {\delta}{\delta q(x)}}\right)\Gamma,
\en 
which can be used to construct charges or simplify calculations in applying consistency 
conditions to all orders in $\hbar$.  For example, we can obtain
\eqa
\partial^\mu{\hat{D}}_\mu&=&\lim_{\lambda(x)\to\lambda}
\left(\beta_\lambda{\frac {\delta}{\delta \lambda(x)}}-\half \gamma\phi(x)
{\frac {\delta}{\delta \phi(x)}}+\half \alpha_m{\frac {\delta}{\delta q(x)}}\right),\\
\partial^\mu \hat{K}_{\mu\nu}&=&
\lim_{\lambda(x)\to\lambda} 2x_\nu
\left(\beta_\lambda{\frac {\delta}{\delta \lambda(x)}}-\half \gamma\phi(x)
{\frac {\delta}{\delta \phi(x)}}+\half \alpha_m{\frac {\delta}{\delta q(x)}}\right).
\ena
where ${\hat{D}}_\mu$ is the current operator for the dilatation transformation 
and $\hat{K}_{\mu\nu}$ is the current operator for the special conformal transformation,
see Appendix C. Hence it is interesting to calculate the conformal transformations
with two external fields $\lambda(x)$ and $q(x)$.

Thirdly, the conformal transformations of the S-matrix with local coupling have been 
calculated in the BPHZ renormalization scheme. Moreover, the S-matrix 
operator with local coupling is a basic object in the  {Epstein}--{Glaser} 
scheme. This means that calculating its conformal transformations is an independent topic,
see \cite{grig, aste}. 
But it is not easy to solve in the  {Epstein}--{Glaser} scheme. Here, it 
can be obtained by direct calculation, although a lot of delicate things lie behind all this. 
It may give some insights into a similar study within the  {Epstein}--{Glaser} scheme.

Lastly, since three methods to describe the conformal transformations of 
the S-matrix are proposed in the paper, they are expected to be also applied 
to fermionic field theories, gauge field theories and supersymmetrical field theories, 
see \cite{christian, elisabeth0}. They will not be much affected by careful 
treatment with  spin dependences in the fermionic field 
theories, gauge fixings in gauge field theories and algebraic  constraints from 
the  {Slavnov}--Taylor identities in the supersymmetrical field theories.


\section{acknowledgements}

I am indebted to Klaus Sibold for initial common work, helpful discussions and 
critical readings on the manuscript. I thank Xiao Yuan Li for helpful comments. 
I would like to thank Christoph Dehne, Markus Roth and Christian Rupp for helpful 
discussions. I would like to thank IHES for its hospitality and thank Dirk Kreimer
for helpful comments.

The DFG is acknowledged for financial support.



\appendix

\section{The cancellation of on-shell poles in
$\Delta_d^\prime\cdot S_n$}

In the on-shell limit, it seems that ${\Delta}^\prime_d\cdot S_{A,n}$ represented
by
{\setlength\arraycolsep{2pt}
\eqa
&&{\Delta}^\prime_d\cdot S_{A,n}=\fur{(i r^{-1/2})}^n\sum_{l=1}^n 2m^2\prodl \,G_{n}\nonumber \\
&& +\fur(i r^{-1/2})^n {\frac {i} {\hbar} }\alpha_m \prodn\Del{d}\cdot G_{n},
\ena}
has on-shell poles like ${\frac {1} {p_i^2-m^2}}$, $p_i$ being the external momenta.
In this section we will prove that the poles of this type do not exist. 
For simplicity, we only treat the S-matrix constructed 
from the connected  {Green} function. 
We denote ${\Delta}^\prime_d\cdot S_{A,n}|_P$ in the momentum space,
{\setlength\arraycolsep{2pt}
\eqa
&&{\Delta}^\prime_d\cdot S_{A,n}|_P=(-1)^n{(i r^{-1/2})}^n(2\pi)^4\delta^4(\sum_{l=1}^n p_l)\nonumber \\
&&\times\{\prod_{i=1}^n(p_i^2-m^2){\frac i \hbar}\alpha_m\Del{d}\cdot \widetilde{G}_{n}-
\sum_{l=1}^n 2m^2\prod_{i=1,i\neq l}^n(p_i^2-m^2)\widetilde{G}_{n}\}|_P.
\ena}

Applying the  {Legendre} transformation, we expand $G_n$ and $\Delta_d\cdot G_n$
respectively, 
{\setlength\arraycolsep{2pt} \eqa
\label{gnn}
G_n&=&{\frac {i} {\hbar}}\int\, \underbrace{G_2\,G_2\cdots G_2}_n\,\Gamma_n+\cdots, \\
\label{dmgnn}
\Delta_d\cdot G_n&=&\int\,
\underbrace{G_2\,G_2\cdots G_2}_n\,\Delta_d\cdot \Gamma_n+\,n{\frac {i} {\hbar}}
\int\, (\Delta_d\cdot G_2)\,\underbrace{G_2\cdots G_2}_{n-1}\,\Gamma_n+\cdots
\ena}
where the symbol $\int$ denotes integration over multi-variables and the symbol
$\cdots$ denotes other unwritten terms which do not affect our proof.
Then we only have to prove that the expression in momentum space, given by
{\setlength\arraycolsep{2pt}
\eqa
\label{specialformulation}
&&(-1)^n{(i r^{-1/2})}^n(2\pi)^4\delta^4(\sum_{l=1}^n p_l)\nonumber \\
&&\times \left\{{\frac i \hbar}\alpha_m (p_1^2-m^2)\Del{d}\cdot
\widetilde{\Delta}(p_1,-p_1)
-2m^2\widetilde{\Delta}(p_1)\right\} 
\nonumber \\
&&\times {\frac i \hbar} \prod_{i=2}^n(p_i^2-m^2)
 \widetilde{\Delta}(p_2)\cdots\widetilde{\Delta}(p_n)\widetilde{\Gamma}_n|_P,
\ena}
has no on-shell pole at $p_1^2=m^2$,
in which the symbol $\widetilde{\Delta}$ stands for the full propagator
(the two-point connected  {Green} function) in the momentum space. 
The employed procedure has an obvious
diagrammatic representation, see FIG. 1. This figure shows the skeleton expansion
of the  {Green} functions. The empty circle denotes the full propagator $G_2$; 
the shaded circle denotes the insertion $\Delta_d\cdot\Gamma_2$; the hatched circle
denotes the kernel $K_n$ which is either the 1PI  {Green} function $\Gamma_n$
or the product of several 1PI  {Green} functions. The numbers $1$, $2$, $\cdots$,
$n$ enumerate all the external lines.

\begin{figure}[htbp]
{\includegraphics[bb=144 288 468 594]{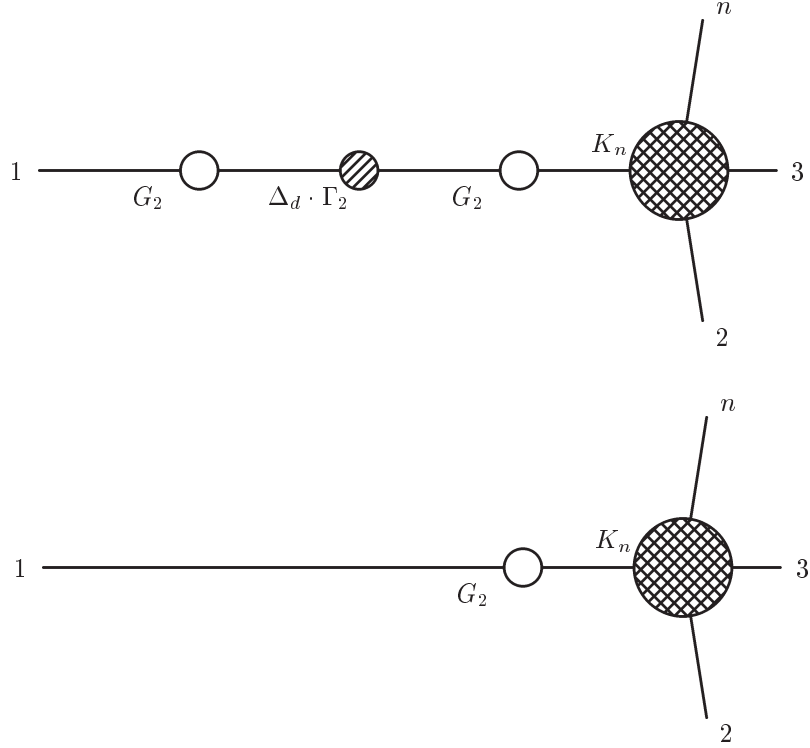}}
\caption{Cancellation of on-shell poles in the dilatation transformation} 
\end{figure}


Before completing the proof, two things have to be prepared. We have realized
{\setlength\arraycolsep{2pt} \eqa
2(1-p^2\partial_{p^2})\widetilde{\Gamma}_{2}{(p,-p)}+
(\beta_{\lambda}\partial_{\lambda}-\gamma)
 \widetilde{\Gamma}_{2}{(p,-p)} \nonumber\\
 =\alpha_m \Del{d}\cdot\widetilde{\Gamma}_{2}{(p,-p)},\ena}
from which the relation between the residue $r$ and the insertion into the 1PI
two-point  {Green} function $\Del{d}\cdot\widetilde{\Gamma}_{2}$ is derived to be
\eq {\frac {-2m^2}
{r}}=\alpha_m\Delta_d\cdot\widetilde{\Gamma}_{2}|_{p^2=m^2}, \en
which is consistent with the normalization conditions
(\ref{normalization}). 
The other crucial point is that the residue $r$ can
be determined in the following way, \eq \lim_{p^2\to
m^2}(p^2-m^2)\widetilde{\Delta}(p)=i\hbar\,r, \en which is
equivalent to  ${\frac 1 r}$$=$
$\partial_{p^2}\,\widetilde{\Gamma}_2(p,-p)|_{p^2=m^2}$.

Now, we can show that the expression (\ref{specialformulation}) is zero by applying
the relation between $\Delta_d\cdot G_2$ and $\Delta_d\cdot \Gamma_2$,
 \eq
\Delta_d\cdot G_2(x,z)=\int {\rm d}^4 z_1
\int {\rm d}^4 z_2\, G_2(x,z_1)G_2(z,z_2) \Delta_d\cdot \Gamma_2(z_1,z_2).
\en
The proof means that
${\Delta}^\prime_d\cdot S_{n}$ only includes the
contributions from the local integral insertion $\Delta_d$ into the internal 
propagators
of the  {Green} function $G_n$, and hence the amputation of the external propagators 
in ${\Delta}^\prime_d\cdot S_{n}$ is well-defined.

One remark is stated on our proof. It is based on 
the  {Legendre} transformation, the skeleton expansion 
of the  {Green} function, the definition of the residue $r$ and 
the on-shell renormalization conditions. So, it is independent of choices 
of both regularization 
schemes and renormalization procedures.

\section{On the on-shell poles in $\alpha\Delta_k^\prime\cdot S_{A,n}\mid_P$}

Whether $\alpha\Delta_k^\prime\cdot S_{A,n}$ contains the on-shell poles or 
not is a serious
problem in calculating the special conformal transformation of the S-matrix. 
If this was so, 
the amputation of external propagators cannot be well-defined. In the following, we 
will try to gain some insights. With  $\alpha\Delta_k^\prime\cdot S_{A,n}$ given by   
\eq
\alpha\Delta_k^\prime\cdot S_{A,n}:= 2m^2\,
\sum_{l=1}^n\,\F^A_n(x,\check{x}_l;p)  (2\alpha x_l)\, G_n +{\frac i \hbar}
\alpha_m\,\F^A_n(x;p) \alpha\Delta_k\cdot G_n,
\en
it is necessary to judge whether the sum of the two terms
\eq
\label{sop}
{\frac i \hbar}  \int {\rm d}^4 x_1 e^{i\,p_1 x_1} \left(2\,m^2(2\alpha x_1)
 G_2(x_1,y_1)+ {\frac i \hbar}\alpha_m (\Box_{x_1}+m^2)\,\alpha\Delta_k\cdot G_2(x_1,y_1)
 \right) 
\en
vanishes in the on-shell limit. The strategy is the same as in the case of 
the dilatation
transformation. Its diagrammatic interpretation is also shown in the FIG. 1, 
except that the symbol 
$\Delta_d\cdot \Gamma_2$ is changed to $\alpha\Delta_k\cdot \Gamma_2$.

By means of the  {Fourier} transformation, 
$\alpha\Delta_k\cdot \Gamma_2(x_1,x_2)$ can be represented by
\eqa
&&\alpha\Delta_k\cdot \Gamma_2(x_1,x_2)=\int{\frac {{\rm d}^4 p_1} {(2\pi)^4}} 
e^{-ip_1(x_1-x_2)} (2\alpha x_2) [\half \phi^2(0)]_2\cdot \widetilde{\Gamma}_2(p_1,-p_1) 
\nonumber\\
&&+\int{\frac {{\rm d}^4 p_1} {(2\pi)^4}} e^{-ip_1(x_1-x_2)}
(-2i\alpha{\frac \partial {\partial p }})[\half\phi^2(p)]_2\cdot
\widetilde{\Gamma}_2(p_1,-p-p_1)\mid_{p=0},
\ena
so the $\alpha\Delta_k\cdot G_2(x_1,y_1)$ is denoted by
\eqa
\alpha\Delta_k\cdot G_2(x_1,y_1)&=&\int{\frac {{\rm d}^4 p_1} {(2\pi)^4}} 
{\frac {i\hbar\, e^{-ip_1 x_1}} {\widetilde{\Gamma}_2(p_1,-p_1)}}
[\half \phi^2(0)]_2\cdot \widetilde{\Gamma}_2(p_1,-p_1) \nonumber\\
&&\times (-2i\alpha{\frac \partial {\partial p_1 }}) 
{\frac {i\hbar\, e^{ip_1 y_1}} {\widetilde{\Gamma}_2(p_1,-p_1)}}
\nonumber\\
&&+\int{\frac {{\rm d}^4 p_1} {(2\pi)^4}}{\frac {i\hbar\, e^{-ip_1 x_1}} 
{\widetilde{\Gamma}_2(p_1,-p_1)}}
{\frac {i\hbar\, e^{ip_1 y_1}} {\widetilde{\Gamma}_2(p_1,-p_1)}} \nonumber\\ 
&&\times (-2i\alpha{\frac \partial {\partial p }})[\half\phi^2(p)]_2\cdot
\widetilde{\Gamma}_2(p_1,-p-p_1)\mid_{p=0}.
\ena

Hence, in $\alpha\Delta_k^\prime\cdot S_{A,n}$, the term $\mathcal{KP}$
containing all possible on-shell poles is given by
\eqa
\label{kp}
&&\mathcal{KP}:=(\,i\,{\hbar}^{n-1}\alpha_m\,r^{{\frac 1 2} n})
(2i\alpha{\frac \partial {\partial p }})
\nonumber\\
&&\left(\sum_{l=1}^n\,{\frac  {[\half\phi^2(p)]_2\cdot \widetilde{\Gamma}_2(p_l,-p-p_l)}
 {\widetilde{\Gamma}_2(p_l,-p_l)}}\right)_{p=0}
 \, K_n(p_1,\cdots,p_n)\mid_{p_l^2=m^2}.
\ena
Formally taking the massless limit, $\mathcal{KP}$ will vanish because of 
the vanishing $\alpha_m$.

In the following, for example, the insertion 
$[\half\phi^2(p)]_2\cdot \widetilde{\Gamma}_2(p_1,-p-p_1)$ is considered.
 Due to the  {Lorentz} invariance and the subtraction scheme 
in the BPHZ renormalization procedure, it is obtained from
\eq
[\half\phi^2(p)]_2\cdot \widetilde{\Gamma}_2(p_1,-p-p_1)
=R(p^2,p_1^2,m^2, p\cdot p_1)-R(0,0,m^2, 0),
\en
where the symbol $R$ stands for the term without subdivergences.
So, the problem changes to the one whether the derivative
\eq 
{\frac {\partial R(0,m^2,p\cdot p_1)}  { \partial p\cdot p_1 }}\mid_{p=0}
\en
vanishes or not, which cannot be exactly solved in a general sense 
at least in a massive scalar field theory. Here it will be calculated 
up to two-loop.

Up to order in $\hbar$, there are two non-vanishing  {Feynman} integrals. 
The two corresponding  {Feynman} diagrams are illustrated in FIG.2.
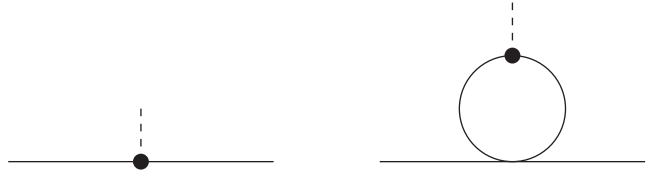
\begin{figure}[htbp]
\begin{center}
\begin{picture}(240,40)(-25,-5)
\Vertex(50,20){3}
\Line(0,20)(100,20)
\DashLine(50,20)(50,40){3}
\Line(140,20)(240,20)
\Vertex(190,60){3}
\DashLine(190,60)(190,80){3}
\CArc(190,40)(20,0,360)
\end{picture}
\caption{Contributions to $\mathcal{KP}$ up to order of $\hbar$}
\end{center}
\end{figure}
The first diagram is the
tree approximation giving the constant value and will vanish 
when taking the derivative. 
The second  {Feynman} integral has the form
\eq
{\frac {i\hbar\lambda} 2}{\frac {\rm d^4 k} {(2\,\pi)^4}}(D_k\,D_{p+k}-D_k^2),
\en  
where the symbol $D_k$ and $D_{p+k}$ are denoted by
\eq
D_k={\frac 1 {k^2-m^2}}, \qquad D_{p+k}={\frac 1 {(p+k)^2-m^2}}.
\en
Applying the derivation action ${\frac \partial {\partial p}}$ then setting $p$ zero, 
the second  {Feynman} diagram still contributes zero. 

In order of $\hbar^2$, there are two types of diagrams, given in FIG.3 and FIG.4. 
First,
consider FIG.3. The first diagram is similar to the second one in FIG. 2, but
involves the counterterm as the interaction vertex, 
then also giving a vanishing result. 
The second  {Feynman} diagram is the scoop one including one tadpole, and
thus giving no contribution. The third one has the  {Feynman} integral with
\eq
{\frac {(i\hbar\lambda)^2} 4}{\frac {\rm d^4 k_1} {(2\,\pi)^4}}
{\frac {\rm d^4 k_2} {(2\,\pi)^4}}
   (D_{k_1}\,D_{p+{k_1}}-D^2_{k_1})(D_{k_2}\,D_{p+{k_2}}-D^2_{k_2}),
\en
which also vanishes in the calculation of ${\mathcal KP}$. 
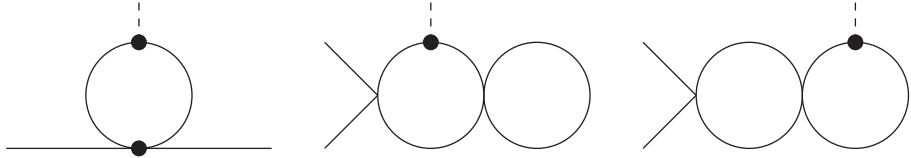
\begin{figure}[htbp]
\begin{center}
\begin{picture}(340,40)(-25,-10)
\Line(0,20)(100,20)
\Vertex(50,60){3}
\Vertex(50,20){3}
\DashLine(50,60)(50,75){3}
\CArc(50,40)(20,0,360)
\Line(120,20)(140,40)
\Line(120,60)(140,40)
\Vertex(160,60){3}
\DashLine(160,60)(160,75){3}
\CArc(160,40)(20,0,360)
\CArc(200,40)(20,0,360)
\Line(240,20)(260,40)
\Line(240,60)(260,40)
\Vertex(320,60){3}
\DashLine(320,60)(320,75){3}
\CArc(280,40)(20,0,360)
\CArc(320,40)(20,0,360)
\end{picture}
\caption{Vanishing contributions to $\mathcal{KP}$ in order of $\hbar^2$ }
\end{center}
\end{figure}

Consider the sunset  {Feynman} diagram in FIG.4.
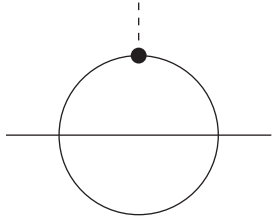
\begin{figure}[htbp]
\begin{center}
\begin{picture}(100,40)(-20,2)
\Vertex(50,90){3}
\Line(0,60)(100,60)
\DashLine(50,90)(50,110){3}
\CArc(50,60)(30,0,360)
\end{picture}
\caption{Non-vanishing contribution to $\mathcal{KP}$ in order of $\hbar^2$}
\end{center}
\end{figure}
The relevant two-loop calculation is carried out to obtain
\eqa
&&(2i\alpha{\frac \partial {\partial p }})
\left(\sum_{l=1}^n\,{\frac  {[\half\phi^2(p)]_2\cdot \widetilde{\Gamma}_2(p_l,-p-p_l)}
{\widetilde{\Gamma}_2(p_l,-p_l)}}\right)^{(\leq 2)}\mid_{p=0,~p_l^2=m^2}\nonumber\\
&&={\frac 1 {2!(4\pi)^4}}(1-{\frac {\pi^2} {12} }){\frac {\hbar^2\lambda^2}{m^2}}
\sum_{l=1}^n\,{\frac {2i\alpha\, p_{l}} {p_l^2-m^2}}\mid_{p_l^2=m^2}.     
\ena

Therefore, in our case, the term 
${\mathcal KP}$ is approximated by
\eq
{\mathcal KP}={\mathcal O}(\hbar^2).
\en
It suggests that 
in the massive scalar field theory $\alpha\Delta_k^\prime\cdot S_{A,n}\mid_{P}$ 
has on-shell
poles. In order to make ${\mathcal KP}$ vanish, there are at least two ways out. 
The first one is that
complicated field theories have to be involved, such as supersymmetrical
field theories, see \cite{white0, white1}. The second one suggests to redefine 
the special conformal transformation of the off-shell S-matrix element $S_{A,n}$ 
in the on-shell limit
\eq
\alpha\mathcal{W}^K S_{A,n}\mid_{P}
\en
according to the equation (\ref{spedeoff})
\eq
\alpha\mathcal{W}^K S_{n}=\alpha\mathcal{W}^K S_{A,n}\mid_P-
4\,i\,m^2\sum_{l=1}^n\alpha\mathcal{P}^k_l S^{\prime}_{A,n}\mid_P,
\en 
since $\alpha\mathcal{W}^K S_{n}$ has to be finite in a physical sense.


\section{Current constructions and charge constructions }

As a matter of fact, the \sm used in our work is defined in the  LSZ
reduction procedure. With this approach, the information
on the operator formalism can be recovered,  such as  current constructions,
charge constructions and  quantum transformations of quantum fields. 

For simplification, the moment constructions of the local  {Ward}
identity operators can be chosen as follows:
\eqa
\widetilde{\mathbf{w}}_{\mu}^T(x)&=&
\partial_{\mu}\phi(x){\frac{\delta}{\delta\phi(x)}}-
{\frac {1} {4}}\partial_{\mu}\left(\phi(x){\frac{\delta}{\delta\phi(x)}}\right),\\
\widetilde{\mathbf{w}}_{\mn}^L(x)&=&x_{\mu}\widetilde{\mathbf{w}}_{\nu}^T(x)-
 x_{\nu}\widetilde{\mathbf{w}}_{\mu}^T(x),\\
\widetilde{\mathbf{w}}^D(x)&=&x^{\mu}\widetilde{\mathbf{w}}_{\mu}^T(x),\\
\widetilde{\mathbf{w}}^K_{\nu}(x)&=&
(2x^{\mu}x_{\nu}-\eta^{\mu}_{\nu}x^2)\widetilde{\mathbf{w}}_{\mu}^T(x).
\ena

Applying the \QAP in the BPHZ renormalization procedure, the energy-momentum 
tensor $T_{\mn}$  can be calculated from
\eq
\widetilde{\mathbf{w}}_{\mu}^T(x)\cdot\Gamma=-[\partial^{\nu}\,T_{\mn}(x)]\cdot\Gamma.
\en
Then the breaking of the conformal invariance are controlled by the insertion of 
the trace of the energy-momentum tensor, namely,
\eqa
\widetilde{\mathbf{w}}^D\Gamma&=&-\partial^{\nu}\left[{D_{\nu}}\right]\cdot\Gamma
    +\left[T^{\nu}_{\nu}\right]\cdot\Gamma,\\
\widetilde{\mathbf{w}}^K_\nu\Gamma&=&
-\partial^{\mu}\left[{K_{\mu\nu}}\right]\cdot\Gamma
    +2x_\nu\left[T^{\mu}_{\mu}\right]\cdot\Gamma,
\ena
where the current $D_{\nu}$ is $x^{\mu}T_{\mn}$  for the dilatation transformation  
and the current $K_{\mu\nu}$ is 
$(2x_{\nu}x^{\zeta}-\eta_{\nu}^{\zeta} x^2)T_{\mu\zeta}$ 
for the special conformal transformation.

Define the local  {Ward} identity for the space-time translation in the 
generating functional $Z[J]$ by
\eqa
\widetilde{\mathbf{w}}_{\mu}^T(x)Z[J]&:=&
\left(J(x)\partial_{\mu}{\frac {\delta }{ \delta J(x)}}-
{\frac 1 4}\partial_\mu(J(x) {\frac {\delta }{ \delta J(x)}})
\right)Z[J] \nonumber\\
&=&{\frac i \hbar}\partial^{\nu}[T_{\mn}]\cdot Z[J].
\ena
It can be realized in the  {Green} function via
\eqa
\widetilde{\mathbf{w}}_{\mu}^T(x)G_n&=&\sum_{l=1}^n\delta(x-x_l)\partial^x_{\mu}
G_n(x,x_1,\cdots, \check{x}_l,\cdots, x_n)\nonumber\\
&&-{\frac 1 4}\sum_{l=1}^n\partial^x_{\mu}\left(\delta(x-x_l)
G_n(x,x_1,\cdots, \check{x}_l,\cdots, x_n)\right)
\nonumber\\
&=&{\frac i \hbar}\partial^{\nu}[T_{\mn}]\cdot G_n(x_1,\cdots, x_n),
\ena
where
$\check{x}_l$ indicates that $\check{x}_l$ is missing in the string of variables.
Furthermore, the local  {Ward} identity for space-time translations
in the momentum space is given by
\eqa
\label{lwfst}
\widetilde{\mathbf{w}}_{\mu}^T(p)G_n&=&-i\sum_{l=1}^n
\left((p_\mu+p_{l,\mu})-{\frac 1 4} p_\mu \right) G_n(p+p_l,p_1,\cdots,\check{p}_l,\cdots,p_n)
\nonumber\\
&=&{\frac i \hbar}(-ip^{\nu})[T_{\mn}(p)]\cdot G_n(p_1,\cdots,\cdots,p_n),
\ena
which can be transfered into the form
\eqa
&&{\frac i \hbar}(-ip^{\nu})[T_{\mn}(p)]\cdot S_n(p_1,\cdots,\cdots,p_n)\nonumber\\
&&=-i\sum_{l=1}^n \left((p_\mu+p_{l,\mu})-{\frac 1 4} p_\mu \right)
(-ir^{-\half})^n \prod_{j=1}^n(p_j^2-m^2) \nonumber\\
&&\times G_n(p+p_l,p_1,\cdots,\check{p}_l,\cdots,p_n)|_P.
\ena
In the on-shell limit and in case of $p$ being nonzero 
and the right hand side being zero,
the conservation of the  energy-momentum tensor is obtained by
\eq
p^{\nu}\hat{T}_{\mn}(p)=0,
\en
which is given in coordinate space by
\eq
\partial^{\nu}\hat{T}_{\mn}=0.
\en
Hence the four-momentum charge $\hat{P}_{\mu}$ is defined as
\eq
\hat{P}_{\mu}:=\int\,{\rm d}^3 x \,\hat{T}_{\mu 0}.
\en
Here all involved operators are defined in the asymptotic  {Hilbert}
space $\mathcal{H}$ satisfying
\eq
\mathcal{H}=\mathcal{H}_{\rm in}=\mathcal{H}_{\rm out}.
\en

Similarly, for the other conformal transformations, the corresponding expressions
can be also set up
\eqa
\partial^a \hat{M}_{\mn a}&=&0,\nonumber\\
\partial^{\mu} \hat{D}_\mu&=&\hat{T}^\nu_\nu,\nonumber\\
\partial^{\mu} \hat{K}_{\mu\nu}&=&2x_\nu\hat{T}^\mu_\mu,
\ena
where $\hat{M}_{\mn a}$, $\hat{D}_\mu$, and $\hat{K}_{\nu\mu}$ are
respectively given by
\eq
\hat{M}_{\mn a}= x_\mu\hat{T}_{\nu a}-x_\nu\hat{T}_{\mu a},\qquad
\hat{D}_\mu = x^\nu\hat{T}_{\mn},\qquad
\hat{K}_{\nu\mu}=(2x_\nu x^\zeta-\eta_{\nu}^\zeta x^2) \hat{T}_{\mu\zeta}.
\en
Then the charge $M_{\mn}$ for the  {Lorentz} rotations is denoted by 
$\int\,{\rm d}^3 x \hat{M}_{\mn 0} \,$.  But the 
charges for both the dilatation transformation and for the special conformal
transformation cannot easily be found out. 

To derive the quantum transformations of the quantum field operator $\hat{\Phi}$, 
the LSZ reduction procedure can be applied 
on the both sides of the local  {Ward} identity (\ref{lwfst}), namely
\eqa
&&{\frac i \hbar}(-ir^{-\half})^n
\prod_{j=2}^n(p_j^2-m^2)(-ip^{\nu})[T_{\mn}(p)]\cdot G_n(p_1,\cdots,\cdots,p_n)|_P
\nonumber\\
&&=-i\sum_{l=1}^n \left((p_\mu+p_{l,\mu})-{\frac 1 4} p_\mu \right)
(-ir^{-\half})^n \prod_{j=2}^n(p_j^2-m^2) \nonumber\\
&&\times\,G_n(p+p_l,p_1,\cdots,\check{p}_l,\cdots,p_n)|_P.
\ena
In the on-shell limit, the above formalism is related to
\eq
-i\left((p_\mu+p_{l,\mu})-{\frac 1 4} p_\mu \right)\hat{\Phi}(p+p_1)
={\frac i \hbar}(-ip^{\nu})\mathcal{T}\left(\hat{T}_{\mn}(p)\hat{\Phi}(p_1)\right),
\en
where the symbol $\mathcal{T}$ denotes the time-ordering defined in the coordinate space.
Then the action of the local  {Ward} identity operator for space-time
translations on the quantum field operator $\hat{\Phi}(x)$ in coordinate space
is given by
\eqa
\label{sttransformation}
\widetilde{\mathbf{w}}_{\mu}^T(x)\hat{\Phi}(x_1)&:=&\partial_\mu^x\hat{\Phi}(x)\delta^4(x-x_1)-
{\frac 1 4}\partial_\mu^x(\hat\Phi(x)\delta^4(x-x_1))\nonumber\\
&=&{\frac i \hbar}\partial^\nu\mathcal{T}\left(\hat{T}_{\mn}(x)\hat{\Phi}(x_1)\right).
\ena

The similar equations for the other conformal transformations
can be also obtained by
\eqa
\label{lotransformation}
(x_\mu\widetilde{\mathbf{w}}_{\nu}^T(x)-x_\nu\widetilde{\mathbf{w}}_{\mu}^T(x))\hat{\Phi}(x_1)
&=&{\frac i \hbar}\partial^a\mathcal{T}\left(\hat{M}_{\mn a}(x)\hat{\Phi}(x_1)\right),\\
x^{\mu}\widetilde{\mathbf{w}}_{\mu}^T(x)\hat{\Phi}(x_1)&=&
{\frac i \hbar}\partial^\mu\mathcal{T}\left(\hat{D}_{\mu}(x)\hat{\Phi}(x_1)\right)
\nonumber\\
&&-{\frac i \hbar}\mathcal{T}\left(\hat{T}^\nu_{\nu}(x)\hat{\Phi}(x_1)\right),\\
(2x_\nu x^\mu-\eta_{\nu}^\mu x^2)\widetilde{\mathbf{w}}_{\mu}^T(x)\hat{\Phi}(x_1)
&=& {\frac i \hbar}\partial^\mu\mathcal{T}\left(\hat{K}_{\mu\nu}(x)\hat{\Phi}(x_1)\right)
\nonumber\\
&&-{\frac i \hbar}2x_\nu\mathcal{T}\left(\hat{T}^\mu_{\mu}(x)\hat{\Phi}(x_1)\right).
\ena
The quantum transformations of the quantum field $\hat{\Phi}$ 
 for space-time translations and  {Lorentz} rotations
are obtained by integrating
\eq
\int^{x^0+\eps}_{x^0-\eps}{\rm d}x_1^0\int {\rm d}^3x_1 
\en
on both sides of the above
equations (\ref{sttransformation}) and (\ref{lotransformation}), namely
\eqa
\delta^T_\mu\,\hat{\Phi}&:=&\partial_\mu\hat{\Phi}={\frac i \hbar}[\hat{P}_\mu,\hat{\Phi}],\nonumber\\
\delta^L_{\mn}\,\hat{\Phi}&:=&(x_\mu\partial_\nu-x_\nu\partial_\mu)\hat{\Phi}=
{\frac i \hbar}[\hat{M}_{\mn},\hat{\Phi}].
\ena

In the cases of the dilatation transformation and the special
conformal transformation, the quantum 
transformations in the free(or asymptotic free) field theory are constructed by
\eqa
\delta^D\hat{\phi}_{\rm in}(x)&:=&(1+x\partial_x)\hat{\phi}_{\rm in}(x)=
{\frac i \hbar}[\hat{D},\hat{\phi}_{\rm in}(x)],
\nonumber\\
\delta^K_\nu \hat{\phi}_{\rm in}(x)&:=&
\left((2x_\nu x^\mu-\eta_\nu^\mu x^2)\partial_\mu+2x_\nu\right)\hat{\phi}_{\rm in}(x)
={\frac i \hbar}[\hat{K}_\nu,\hat{\phi}_{\rm in}(x)],
\ena
where $\hat{D}$ is the charge for the dilatation transformation and 
$\hat{K}_\nu$ is the charge for the special conformal transformation.




\end{document}